\documentclass[aps,twocolumn,amsfonts,amssymb,prb,superscriptaddress,10pt]{revtex4-2}

\usepackage{graphicx}
\usepackage{xcolor}
\usepackage{amsmath}
\usepackage{hyperref}
\usepackage{orcidlink}
\hypersetup{colorlinks=true,citecolor=blue,linkcolor=blue,urlcolor=blue}

\allowdisplaybreaks 

\begin{document}

\title{Structure and relaxation of long-lived odd-parity modes in two-dimensional Fermi liquids}

\author{Stan Boeve}
\affiliation{Department of Physics, Gothenburg University, 41296 Gothenburg, Sweden}

\author{Johannes Hofmann\,\orcidlink{0000-0002-0667-2452}}
\email{johannes.hofmann@physics.gu.se}
\affiliation{Department of Physics, Gothenburg University, 41296 Gothenburg, Sweden}
\affiliation{Nordita, Stockholm University and KTH Royal Institute of Technology, 10691 Stockholm, Sweden}

\date{\today}

\begin{abstract}
In two-dimensional Fermi liquids, odd-parity deformations of the Fermi surface relax anomalously slowly at a rate $\sim T^4$, where $T$ is the temperature. However, the leading-logarithmic prediction for the magnitude of this rate $\sim m^4 \ln m$, where $m$ is an angular harmonic index, exceeds exact-diagonalization results and the $\ln m$ scaling is not observed. We resolve this discrepancy by evaluating the 2D Fermi-liquid collision integral for general regular isotropic potentials to leading order in $T/T_F$, including the nonlogarithmic term beyond the  $\ln(T_F/T)$ term. We find that the local temperature modulation of the odd mode, which we call the thermal dressing, suppresses relaxation by collinear scattering, with residual relaxation set by large-angle scattering. Such large-angle processes give corrections of order $\ln m/\ln(T_F/T)$ that are not small and must be resummed. We benchmark our results against exact-diagonalization calculations for constant interactions and find agreement over a range $m^2 T/T_F \lesssim 0.4$, and we provide closed-form results for the slowest odd-mode relaxation rate of a spin-polarized dipolar gas. 
\end{abstract}

\maketitle

\section{Introduction}\label{sec:introduction}

Two-dimensional (2D) scattering kinematics imposes strong constraints on the quasiparticle relaxation in a Fermi liquid at low temperatures. Here, for scattering states confined to the Fermi surface, energy and momentum conservation allow only two types of binary collisions: First, processes that leave initial states unchanged up to relabeling (these describe direct and exchange scattering), and, second, head-on scattering processes, in which two particles with momenta at antipodal points of the Fermi surface scatter into a different antipodal pair. The former do not relax a quasiparticle distribution, while the latter lead to a characteristic relaxation rate $\gamma_{\rm fast} \sim T^2/(\hbar T_F)$ that follows the standard quadratic-in-temperature Fermi-liquid form ($T_F$ is the Fermi temperature and we set $k_B=1$). 

As first pointed out by \citet{gurzhi82}, however, this head-on relaxation is parity-sensitive: Since head-on scattering affects opposite points on the Fermi surface equally, only collective deformations of the quasiparticle distribution that are parity-even will relax in this way. Odd-parity deformations, by contrast, relax by processes with a finite-temperature deviation ${\it O}(T/T_F)$ from the zero-temperature kinematics, which leads to a much slower rate $\gamma_{\rm slow} \sim T^4/(\hbar T_F^3)$~\cite{gurzhi82,gurzhi87,gurzhi93,gurzhi95c}. This scaling was established for deformations that vary over the scale of the Fermi surface, with~\citet{gurzhi93} noting that a different analysis would be required for strongly anisotropic odd-parity deformations. 

%++++++++++++++++++++++++++
\begin{figure}[b]
	\centering
 \scalebox{0.9}{\includegraphics{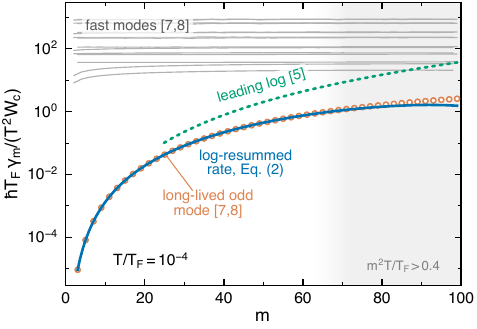}}
 \caption{
 Relaxation spectrum of a 2D Fermi liquid with constant interaction $W_c$ [Eq.~\eqref{eq:constantamplitude}] at $T/T_F=10^{-4}$ shown as a function of the angular harmonic index $m$. Orange circles: slow odd-mode branch. Gray lines: remaining fast modes. Green dotted line: asymptotic prediction of Ref.~\cite{ledwith19}. Blue line: this work, Eq.~\eqref{eq:oddmoderesult} minimized over the relative dressing $\xi$ [Eq.~\eqref{eq:gammamconstoptimal}]. The shaded gray region $m^2T/T_F\gtrsim0.4$ lies outside the region of applicability of the low-temperature expansion.
 }
 \label{fig:1}
\end{figure}
%++++++++++++++++++++++++++

This analysis was supplied in the more recent literature by \citet{ledwith19}, who resolved the angular-harmonic structure of odd modes and obtained $\gamma_{\rm slow} = (4\pi^3/15) T^4/(\hbar T_F^3) W_1 m^4 \ln m$, where $m$ is the angular harmonic index of the Fermi-surface deformation and $W_1$ is the collinear head-on scattering amplitude [cf.~Eq.~\eqref{eq:W1def}]. The $\ln m$ factor comes from the leading $\ln(T_F/T)$ divergence of the collision matrix elements, which arises from collinear small-angle head-on scattering processes, and the result applies for \mbox{$\ln m \gg 1$}~\cite{ledwith19}. Subsequent work found from a supersymmetric mapping a single long-lived odd mode per angular sector~\cite{kryhin23}. Exact diagonalization (ED) studies of the electron collision integral confirm the existence of a band of long-lived odd-parity modes in a wide temperature range $T\lesssim 0.15T_F$, one per angular harmonic sector up to $m\lesssim \sqrt{T_F/T}$, for a range of interaction potentials~\cite{hofmann23,nilsson25,maki25}. These calculations show the $m^4T^4$ scaling at low temperatures, but (i) they do not show the logarithmic $m$-scaling and (ii) the rates are much smaller than predicted in Ref.~\cite{ledwith19}. This is illustrated in Fig.~\ref{fig:1}, which shows the relaxation spectrum for a constant interaction potential obtained in Refs.~\cite{hofmann23,nilsson25} via exact diagonalization at $T/T_F=10^{-4}$ [gray lines], with the lowest odd branch highlighted [orange circles]. The asymptotic prediction by Ref.~\cite{ledwith19} (green dotted line) exceeds the numerical rates for all $m$ in its range of applicability by up to an order of magnitude, and there is no $m$-range where the $m^4\ln m$ scaling form agrees with the numerical rates.

The disagreement arises because the leading-logarithmic result requires two limits that in practice are not reached at the same time. First, the collinear logarithms in $m$  dominate finite terms only for $\ln m \gg 1$. Second, corrections to the leading-logarithmic result, which are of relative order $\ln m/[\ln(T_F/T)/2]$, need to be suppressed, i.e., $\ln m \ll \ln(T_F/T)/2$. The second condition is a much stronger requirement than the condition $m^2 T/T_F \ll 1$ for the validity of the low-temperature expansion. Even if each inequality holds by only a factor of $5$, i.e., $\ln m = 5$ and $\ln(T_F/T)/2 = 5 \ln m$, this would require $T/T_F \lesssim e^{-50}$.

In this work, we compute the low-temperature collision integral in closed analytic form for general regular isotropic interaction potentials, where we include both logarithmic and nonlogarithmic contributions. Throughout we consider a parabolic single-particle dispersion with effective mass $m^*$ and an isotropic regular interaction potential (i.e., potentials that are a smooth function of the transferred momentum and remain finite in the long-wavelength limit). The longest-lived odd mode consists of a rigid Fermi-surface deformation together with a local temperature change, and our closed-form result shows that this modulation allows it to avoid relaxation by collinear scattering, which would otherwise be the dominant relaxation channel. Microscopically, the long-lived odd mode in the $m$th angular sector is a deviation of the quasiparticle distribution from the equilibrium Fermi-Dirac distribution $f_0$ of the form
\begin{align}
 \delta f_m({\bf k}) &\sim \biggl(- \frac{\partial f_0}{\partial \varepsilon}\biggr) \Bigl(1 + \xi \frac{m^2 T}{2 T_F} \frac{\varepsilon-\mu}{T}\Bigr) e^{im\theta} , \label{eq:odddeformation}
\end{align}
where $\theta$ is the polar angle of ${\bf k}$, $\varepsilon$ is the quasiparticle energy, and $\mu$ is the chemical potential. The first term in parentheses is a local shift in the chemical potential (i.e., a rigid displacement of the Fermi surface), and the second term describes a local change in temperature. We call the temperature modulation the thermal dressing, and $\xi$ the relative dressing. The value $\xi=0$ describes a rigid deformation, and the prefactor $m^2 T/(2 T_F)$ is chosen such that $\xi=1$ describes a fully dressed mode that is not relaxed at all by collinear scattering. Within this two-moment variational ansatz, we find that the deformation relaxes with the rate 
\begin{align}
&\gamma_m(\xi) = \frac{\pi^3}{15} \frac{m^4 T^4}{\hbar T_F^3} 
 \biggl\{ 
\underbrace{2 W_1 \bar{\Lambda}_T \bigl(\xi - 1\bigr)^2}_{\rm collinear \ scattering}  \nonumber \\
& +
\underbrace{\int_0^1 \frac{ds}{s} \biggl[W_{\rm tot}(s) U_{m-1}^2(\sqrt{s}) \bigl(\xi - \bar{\xi}(s)\bigr)^2- 2 W_1 \bigl(\xi - 1\bigr)^2\biggr]}_{{\rm large\text{-}angle \ scattering}}
\biggr\} , \label{eq:oddmoderesult}
\end{align}
where $\bar{\Lambda}_T=\Lambda_T + h$ with $\Lambda_T=\ln (T_F/T)$ and $h= \ln 4 - \psi(5) - \zeta'(4)/\zeta(4)=-0.05615$ [Sec.~\ref{sec:twobasis} and Eq.~\eqref{eq:defhr}], and $U_m$ is a Chebyshev polynomial of the second kind. The integration runs over the momentum transfer $q$ in a binary collision with $s = (q/2k_F)^2$. $W_{\rm tot}(s)$ [Eq.~\eqref{eq:Wtotdef}] is an effective scattering amplitude that combines head-on and direct/exchange branches, and $W_1 = W_{\rm tot}(0)/2$ is the collinear scattering amplitude. Equation~\eqref{eq:oddmoderesult} separates the contributions of different microscopic scattering processes to the decay: As $\Lambda_T\to\infty$, the collinear term [first line in Eq.~\eqref{eq:oddmoderesult}] is leading and forces $\xi=1$, for which collinear scattering does not relax the mode. As discussed above, this limit is only approached logarithmically slowly (for a constant interaction, for example, $\xi_3\approx 0.71$ for $m=3$ even at $T/T_F=10^{-4}$, cf.~Fig.~\ref{fig:5}). The small residual relaxation rate is then set by large-angle scattering [second line in Eq.~\eqref{eq:oddmoderesult}]. By Eq.~\eqref{eq:oddmoderesult}, the large-angle term measures the variance around 
\begin{align}
\bar{\xi}(s) &= \frac{T_{m}(\sqrt{s})}{m \sqrt{s} U_{m-1}(\sqrt{s})} ,
\end{align}
where $T_m$ is a Chebyshev polynomial of the first kind. Here, $\bar{\xi}(s)$ is the relative dressing for which collisions with momentum transfer $s$ do not relax the mode. This dressing cannot be chosen independent of $s$ except for the special case $m=1$, for which $\bar{\xi}(s) \equiv1$ and \mbox{$\gamma_{m=1} = 0$}: For $\xi=1$, Eq.~\eqref{eq:odddeformation} describes a uniform momentum boost since $k/k_F = \sqrt{1 + w T/T_F} \simeq 1 + w T/(2T_F)$ with $w=(\varepsilon-\mu)/T$, and momentum is conserved. For general odd $m\geq 3$, however, no fixed choice of $\xi$ eliminates all large-angle processes, and a small nonzero relaxation rate remains.

With increasing $m$, the collinear protection is gradually lost. Formally, the endpoint regions of the integral in Eq.~\eqref{eq:oddmoderesult} contribute terms of order $\ln (m^2)$ that compete with the collinear term $2W_1\bar{\Lambda}_T$ (whereas full dressing $\xi \simeq 1$ requires $\ln(m^2) \ll \bar{\Lambda}_T$). These terms arise from integration regions with momentum transfer $s \gtrsim 1/m^2$, for which $\bar{\xi}(s)$ is no longer close to unity. Hence, these collisions do relax a dressed mode. As $m$ is increased toward $m \simeq \sqrt{T_F/T}$, the optimal dressing decreases toward $\xi\approx 0$ (i.e., an essentially rigid Fermi-surface deformation), which relaxes much more slowly than the fully dressed mode would [cf.~Fig.~\ref{fig:5}(b)]. In this large-$m$ limit, collinear collisions relax the mode but they do not set its dressing. When keeping the thermal dressing fixed at \mbox{$\xi=1$} and expanding for large $\ln m$, we recover the functional form of Ref.~\cite{ledwith19}, with a different coefficient that includes the forward-scattering amplitude $W_0$ in addition to $W_1$. 

The minimum of the relaxation rate in  Eq.~\eqref{eq:oddmoderesult} is shown as a blue solid line in Fig.~\ref{fig:1}, and it agrees well with the exact-diagonalization result over the full $m$-range of the low-temperature description. The analytic results for the collision integral obtained in this work can be used as input parameters in transport calculations beyond simple relaxation-time models. In particular, the companion Letter discusses how the thermal dressing should be detectable in the thermopower~\cite{hofmann26}.

This paper is structured as follows: Section~\ref{sec:kinematics} discusses the phase space for binary scattering at low temperatures and establishes the different kinematic regions that contribute to the collision integral. Section~\ref{sec:asymptoticmatching} evaluates the coefficients of the low-temperature collision matrix~\eqref{eq:structure} for general regular isotropic potentials by asymptotic matching of the collision integral in momentum-transfer variables. The results for leading logarithmic and finite contributions are collected in Sec.~\ref{sec:asymptoticresults}. Subsequent sections use only these results, so a reader mainly interested in the applications of our results can skip Secs.~\ref{sec:matching} and~\ref{sec:matchingderivation}. Section~\ref{sec:eigenmodes} derives the structure of the long-lived odd modes and the closed-form expression~\eqref{eq:oddmoderesult} discussed above. In Sec.~\ref{sec:constant}, we evaluate the collision matrix explicitly for a constant interaction and benchmark against exact-diagonalization results of Refs.~\cite{hofmann23,nilsson25}. As an application of our formalism, Sec.~\ref{sec:dipoles} provides results for the relaxation rates of odd eigenmodes of spin-polarized dipolar fermions, which are relevant for experiments with dipolar quantum gases. Section~\ref{sec:summary} concludes with a summary and outlook. Technical details of our calculations are collected in Appendixes~\ref{app:collisionintegral}--\ref{app:asymptotickernels}, \ref{app:logfermiintegrals}, and~\ref{app:Fermiintegrals}. Appendix~\ref{app:L01} derives the leading off-diagonal collision element for even $m$, which defines a relative even-mode dressing analogous to that in Eq.~\eqref{eq:odddeformation}, and which contributes to the thermopower studied in the companion Letter~\cite{hofmann26}.

\section{Two-particle scattering kinematics}\label{sec:kinematics}

Before evaluating the collision integral, we summarize the kinematics of binary scattering in two dimensions. This kinematics has been described starting from the thermally smeared scattering momenta in Refs.~\cite{laikhtman92,gurzhi82,gurzhi87,gurzhi93,menashe96,ledwith19}, among others. Here, we derive all constraints in a unified way from the total energy of scattering particles.

%++++++++++++++++++++++++++
\begin{figure}
	\centering
 \includegraphics{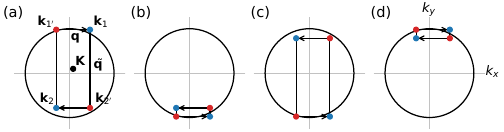}
 \caption{Allowed momentum-space configurations at fixed ${\bf q}$ and $\omega$. Initial momenta ${\bf k}_1$ and ${\bf k}_2$ (final momenta ${\bf k}_{1'}$ and ${\bf k}_{2'}$) are shown in blue (red). We orient ${\bf q}$ along the $x$ axis, and the Fermi surface is shown for comparison. The endpoints of the four scattering momenta enclose a rectangle at all temperatures [Eq.~\eqref{eq:rectangle}]. Energy and momentum conservation constrain only the magnitude and projection along ${\bf q}$, which  leaves four choices for the perpendicular transfer $\tilde{\bf q}$ [Eq.~\eqref{eq:costheta}], corresponding to placing each of the pairs $({\bf k}_1, {\bf k}_{1'})$ and $({\bf k}_2, {\bf k}_{2'})$ in the upper or lower half-plane.
}
 \label{fig:2}
\end{figure}
%++++++++++++++++++++++++++

A two-body scattering process ${\bf k}_1 + {\bf k}_2 \to {\bf k}_{1'} + {\bf k}_{2'}$ is described by initial momenta ${\bf k}_{1}$ and  ${\bf k}_{2}$ and final momenta ${\bf k}_{1'}$ and ${\bf k}_{2'}$. The corresponding single-particle energies are measured with respect to the chemical potential 
\begin{align}
 w_\alpha = \beta \Bigl(\frac{\hbar^2 k_\alpha^2}{2m^*} - \mu\Bigr) \label{eq:defw}
\end{align} 
with $\alpha=1,2,1',2'$ and $\beta=1/T$. We denote the corresponding quasiparticle velocities by $v_\alpha = \hbar k_\alpha/m^*$. Such a scattering process is illustrated in Fig.~\ref{fig:2}, where the circle indicates the Fermi surface, blue (red) points are initial (final) momenta, and the black dot indicates the center of mass ${\bf K} = ({\bf k}_1+{\bf k}_2)/2$ of the scattering momenta. Arrows indicate the momentum transfer from the first initial to the two final states,
\begin{alignat}{2}
{\bf q} &= {\bf k}_{1} - {\bf k}_{1'} = {\bf k}_{2'} - {\bf k}_{2}, &\qquad x &= \frac{q}{2k_F}, \\
\tilde{\bf q} &= {\bf k}_{1} - {\bf k}_{2'} = {\bf k}_{1'} - {\bf k}_{2}, &\qquad \tilde{x} &= \frac{\tilde{q}}{2k_F} ,
\end{alignat}
where the second equality follows from momentum conservation and we also define the dimensionless magnitudes $x$ and $\tilde{x}$ of these vectors. Momentum conservation and energy conservation in a parabolic band imply the well-known geometric relation, valid independent of temperature~\cite{laikhtman92,musser26},
\begin{align}
 {\bf q} \cdot \tilde{\bf q} = 0 , \label{eq:rectangle}
\end{align}
i.e., the endpoints of the scattering momenta form a rectangle [cf.~Fig.~\ref{fig:2}]. In the following, scattering geometries are always pictured such that the vector ${\bf q}$ is oriented along the positive $x$ axis. The transfer vectors ${\bf q}$ and $\tilde{\bf q}$ set the relative energy of the scattering particles as $E_{\rm rel} = (\hbar^2/4m^*) (q^2+\tilde{q}^2)$.

A given momentum transfer $q$ and energy transfer $\hbar \omega = \varepsilon_1-\varepsilon_{1'}$ constrain the polar angles of the initial-state vectors with respect to the vector ${\bf q}$ as
\begin{subequations}\label{eq:costheta}
\begin{align}
\cos \theta_1^\star &= + \frac{q}{2k_1} + \frac{\omega}{v_1 q} , \label{eq:costhetat1} \\
\cos \theta_2^\star &= - \frac{q}{2k_2} + \frac{\omega}{v_2 q} \label{eq:costhetat2} .
\end{align}
Each equation has two solutions with opposite signs for $\sin \theta_1^\star$ and $\sin \theta_2^\star$. The final-state vectors are shifted by ${\bf q}$, and their polar angles
satisfy
\begin{align}
\cos \theta_{1'}^\star &= - \frac{q}{2k_{1'}} + \frac{\omega}{v_{1'} q} , \\
\cos \theta_{2'}^\star &= + \frac{q}{2k_{2'}} + \frac{\omega}{v_{2'} q} .
\end{align}
\end{subequations}
By construction, these vectors are in the same momentum half-plane as ${\bf k}_1$ and ${\bf k}_2$, respectively, i.e., ${\rm sgn} \sin \theta_{1'}^\star = {\rm sgn} \sin \theta_1^\star$ and ${\rm sgn} \sin \theta_{2'}^\star = {\rm sgn} \sin \theta_2^\star$. The four different solutions of Eq.~\eqref{eq:costheta} are illustrated in Fig.~\ref{fig:2}. Geometrically, the energy transfers $\hbar \omega$ and $\hbar \tilde{\omega} = \varepsilon_1 - \varepsilon_{2'}$ between $1$ and $1'$ or $2'$ are proportional to the projection of the center-of-mass momentum ${\bf K}$ onto the transfer vectors:
\begin{align}
\omega &= \frac{\hbar}{m^*} {\bf K} \cdot {\bf q} \\
\tilde{\omega} &= \frac{\hbar}{m^*} {\bf K} \cdot \tilde{\bf q} 
\end{align}
(see~also Sec.~3.6 of Ref.~\cite{gurzhi87}, where $K_0$ and ${\bf K}$ are used instead of $\omega$ and ${\bf q}$). This implies for the magnitude of the center-of-mass momentum
\begin{align}
 \frac{K^2}{k_F^2} = \frac{T^2}{16 T_F^2} \Bigl( \frac{w^2}{x^2} + \frac{\tilde{w}^2}{\tilde{x}^2} \Bigr) , \label{eq:commomentum}
\end{align}
where we introduce the dimensionless energy transfers $w=\beta \hbar \omega$ and $\tilde{w}=\beta\hbar\tilde{\omega}$, which fix the center-of-mass energy $E_{\rm com} = \hbar^2 K^2/m^*$. The total energy $E_{\rm tot} = (\hbar^2/2m^*) (k_1^2 +k_2^2)$ can now be expressed in two ways: In terms of the center-of-mass and relative contributions, and directly from Eq.~\eqref{eq:defw} using $T w_i/T_F = k_i^2/k_F^2 - \mu/E_F$. Equating them gives a constraint equation on the magnitude of the momentum transfers $x$ and $\tilde{x}$,
\begin{align}
x^2 + \tilde{x}^2 + \frac{T^2}{16 T_F^2} \Bigl( \frac{w^2}{x^2} + \frac{\tilde{w}^2}{\tilde{x}^2} \Bigr) = \frac{\mu}{E_F} + \frac{T}{2T_F} \bigl(w_1 + w_2\bigr) . \label{eq:constraint}
\end{align}

%++++++++++++++++++++++++++
\begin{figure}
	\centering
 \includegraphics{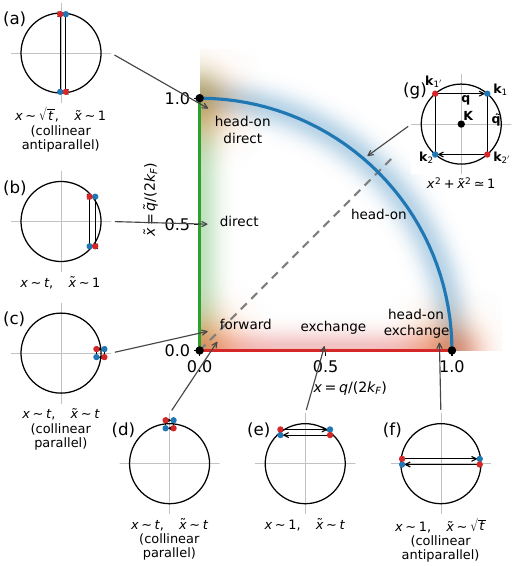}
 \caption{Sketch of the allowed low-temperature scattering processes as a function of the momentum transfers $x$ and $\tilde{x}$. The shaded regions indicate thermal broadening of the allowed momenta. The dashed line indicates the $x \leftrightarrow \tilde{x}$ duality under exchange of final states. Different inset diagrams show typical scattering configurations in the different scattering regions along with the magnitude of the momentum transfer, where we abbreviate $t = T/T_F$.}
 \label{fig:3}
\end{figure}
%++++++++++++++++++++++++++

The discussion so far is completely general and applies to scattering processes at any temperature. Now, at strictly zero temperature, all scattering states are confined to the Fermi surface, such that $\varepsilon_\alpha = \mu = E_F$. The constraint~\eqref{eq:constraint} then implies $x^2 + \tilde{x}^2 = 1$. This describes a head-on scattering process in which an initial state with opposite momenta at the Fermi surface (corresponding to ${\bf K} = 0$) scatters into a final-state head-on configuration with rotated momenta. Such a process is sketched in Fig.~\ref{fig:3}(g). This scattering process is the origin of the odd-even effect in quasiparticle relaxation, as discussed in the introduction.

To capture the asymptotic low-temperature scaling of relaxation rates, we need to describe the scattering kinematics for small nonzero temperatures $T\ll T_F$. Scattering states are now confined to an annulus of width ${\it O}(T)$ in energy around the Fermi energy, such that in our parametrization $w_\alpha = {\it O}(1)$ and $\mu \simeq E_F$. The right-hand side of Eq.~\eqref{eq:constraint} is thus unity at this order. The different low-temperature scattering regimes that are consistent with the constraint~\eqref{eq:constraint} are illustrated in Fig.~\ref{fig:3}, which shows possible scattering momenta in the $x$--$\tilde{x}$ plane, with insets showing typical scattering configurations. These configurations are discussed next.

A head-on branch [blue shaded area in Fig.~\ref{fig:3}, panel (g)] with $x^2+\tilde{x}^2=1$ and $K/k_F={\it O}(T/T_F)$ remains. Additional scattering channels open if either $\tilde{x} = {\it O}(T/T_F)$ or $x = {\it O}(T/T_F)$, in which case the center-of-mass term on the left-hand side of Eq.~\eqref{eq:constraint} is ${\it O}(1)$ and balances the energy equation by an additional center-of-mass contribution. The first case with $x = {\it O}(1)$ and $\tilde{x} = {\it O}(T/T_F)$ [red shaded area in Fig.~\ref{fig:3}, panel (e)] describes a process in which ${\bf k}_{1'} \approx {\bf k}_2$ and ${\bf k}_{2'} \approx {\bf k}_1$, i.e., an exchange scattering that approximately swaps the initial momentum states. The second case with $x = {\it O}(T/T_F)$ and $\tilde{x} = {\it O}(1)$ [green shaded area in Fig.~\ref{fig:3}, panel (b)] describes a process in which scattering momenta are approximately unchanged, ${\bf k}_{1'} \approx {\bf k}_1$ and ${\bf k}_{2'} \approx {\bf k}_2$, which describes direct scattering.

Each branch in Fig.~\ref{fig:3} has two endpoints [orange shaded areas], at which the scattering momenta become collinear. The head-on branch merges with the direct and exchange branches near the points $(x,\tilde{x}) = (0,1)$ [panel (a)] and $(1,0)$ [panel (f)]. These regions describe collinear antiparallel processes with near-zero center-of-mass momentum. By Eq.~\eqref{eq:constraint}, the phase space for scattering is enhanced as $x={\it O}(\sqrt{T/T_F})$ and $\tilde{x}={\it O}(\sqrt{T/T_F})$, respectively, such that $K/k_F = {\it O}(\sqrt{T/T_F})$. The enhanced phase space in these two regions gives rise to the logarithmic singularity in temperature, and it is the contribution of these kinematic regions that was quantified in Ref.~\cite{ledwith19}. An additional endpoint singularity arises for $x,\tilde{x} = {\it O}(T/T_F)$ [panels (c) and (d)]. Here, the center-of-mass momentum lies on the Fermi surface, $K/k_F = 1$, and the process describes forward scattering in which all momenta occupy the same patch on the Fermi surface. 

The definition of exchange and direct scattering is interchangeable upon relabeling of the final states $1'$ and $2'$. Such a swap, referred to as a duality transformation in Ref.~\cite{ledwith19}, interchanges the vectors ${\bf q}$ and $\tilde{\bf q}$ (and thus $x$ and $\tilde{x}$). In Fig.~\ref{fig:3}, such a transformation corresponds to mirroring the kinematic regions along the diagonal [gray dashed line in Fig.~\ref{fig:3}] with a corresponding rotation by $90^\circ$ of the inset diagrams. 

\section{Asymptotic matching of the collision integral}\label{sec:asymptoticmatching}

This section evaluates the collision matrix elements to leading order at low temperature. The main technical result of our work is an evaluation of the linearized collision matrix, i.e., of all components
\begin{align}
 L_{n'n}^{(m)} &= \frac{T^2}{\hbar T_F} 
 \begin{pmatrix}
 \frac{T^2}{T_F^2} \bigl(a_{00} \Lambda_T + b_{00}\bigr) & \frac{T}{T_F} \bigl({\bf a}^T \Lambda_T + {\bf b}^T\bigr) \\
 \frac{T}{T_F} \bigl({\bf a} \Lambda_T + {\bf b}\bigr) & A \Lambda_T + B
 \end{pmatrix} , \label{eq:structure}
\end{align}
written here for odd $m$. Here, $n,n'\geq 0$ label radial basis modes. The anomalous higher-order temperature scaling in the first row and column ($T^4$ and $T^3$ instead of $T^2$) is the odd-even effect. [In the corresponding collision matrix for even $m$, by contrast, all leading entries scale as $T^2/(\hbar T_F)$ or $T^2 \Lambda_T/(\hbar T_F)$ with temperature.] The leading logarithmic terms $a_{00}$, $a_n$, and $A_{n'n}$ are set by collinear scattering. The nonlogarithmic coefficients $b_{00}$, $b_n$, and $B_{n'n}$ depend on large-angle scattering and are set by angle integrals over the interaction. We find that all coefficients factorize into a Fermi average over the radial basis functions that can be evaluated in closed analytic form independent of the interaction potential, and angle integrals that contain all information on the interaction potential and the angular harmonic index $m$. 

The starting point of our analysis is the linearized collision integral
\begin{align}
{\cal L}[h({\bf k})] = - \frac{{\cal J}[\delta f({\bf k})]}{f_0({\bf k}) \bigl(1-f_0({\bf k})\bigr)} , 
\label{eq:defL}
\end{align}
where the deviation from equilibrium is written as $\delta f({\bf k}) = f_0({\bf k}) (1-f_0({\bf k})) h({\bf k})$. The linearized collision integral is a positive semidefinite Hermitian operator with respect to the inner product
\begin{align}
\langle h' | h \rangle &= \frac{1}{N_0 T} \int \frac{d{\bf k}}{(2\pi)^2} \, h^{\prime *}({\bf k}) \, f_0({\bf k}) \bigl(1-f_0({\bf k})\bigr) h({\bf k}) , \label{eq:innerproduct}
\end{align}
where $N_0 = m^*/(2\pi\hbar^2)$ is the density of states per spin. General properties and explicit forms are reviewed in App.~\ref{app:collisionintegral}. Our aim is to describe the relaxational eigenmodes, which are the eigenvectors of the linearized collision integral,
\begin{align}
 {\cal L}[h({\bf k})] &= \gamma h({\bf k}) ,
\end{align}
which describe deviations from equilibrium that decay exponentially with a rate set by the eigenvalue $\gamma$. We expand the perturbation $h({\bf k})$ in a basis of angular components and orthogonal polynomials with respect to the inner product. To this end, we determine the matrix elements
\begin{align}
L_{n'n}^{(m)} &= \langle u_{n'} e^{im\theta}| {\cal L} | u_n e^{im\theta} \rangle ,
\end{align}
where $\theta$ is the angle of momentum in the 2D plane and $u_n$ is a polynomial that depends on the magnitude of the momentum. The radial basis functions are chosen orthonormal with respect to the inner product~\eqref{eq:innerproduct}. Explicit forms and identities are listed in App.~\ref{app:basisfunctions}. The key expression that we evaluate parametrizes the integration in terms of energy and momentum-transfer variables [cf.~App.~\ref{app:collisionintegral}]
\begin{align}
L^{(m)}_{n'n} &= \frac{T^2}{\hbar T_F} 
\bigl\langle I^{(m)}_{n'n}
\bigr\rangle_F 
, \label{eq:collisiondefinition}
\end{align}
where
\begin{align}
\langle G \rangle_F &= \int_{-\beta\mu}^\infty dw_1 \int_{-\beta\mu}^\infty dw_2 \int_{-\infty}^\infty dw \, F_{121'2'} G \label{eq:Fermiintegral}
\end{align}
is an energy average over initial energies and the energy transfer weighted with the Fermi-Dirac functions, 
\begin{align}
    F_{121'2'} = f_0(w_1) f_0(w_2) (1 - f_0(w_1-w)) (1 - f_0(w_2+w)) ,
\end{align}
with a kernel
\begin{align}
I^{(m)}_{n'n} &= \frac{1}{16\pi} \int_0^\infty \frac{dx}{x |\sin \theta_1^\star \sin \theta_2^\star \, |} \, \frac{v_F^2}{v_1 v_2} \nonumber \\
&\quad \times \Bigl[ W(x, \tilde{x}_<) \, K_{n'n}^{(m)+}(x) + W(x, \tilde{x}_>) \, K_{n'n}^{(m)-}(x) \Bigr] \label{eq:matrixelementsimplified}
\end{align}
that includes a dimensionless potential
\begin{align}
W(x, \tilde{x}) &= 
\frac{g_s}{2} \biggl[ | \tilde{V}(2 k_F x) |^2 + | \tilde{V}(2 k_F \tilde{x}) |^2 \nonumber \\
&\quad - \frac{2}{g_s} {\rm Re} \bigl\{ [\tilde{V}(2 k_F x)]^* \tilde{V}(2 k_F \tilde{x}) \bigr\} \biggr] \label{eq:defW} 
\end{align}
with $\tilde{V}({\bf q}) = N_0 V({\bf q})$ and spin degeneracy $g_s$. The first term in brackets in Eq.~\eqref{eq:matrixelementsimplified} describes configurations in which all scattering momenta lie in the same (upper) half-plane, and the second term describes configurations in which initial scattering momenta are on opposite momentum half-planes. Here, $\tilde{x}_<$ ($\tilde{x}_>$) is the root of Eq.~\eqref{eq:constraint} for equal-half-plane (opposite-half-plane) configurations, and we define
\begin{subequations}
\begin{align}
K^{(m)+}_{n'n}(x) &= 2 (C_{n'} C_n + S_{n'}^+ S_n^+) \label{eq:Kp} \\
K^{(m)-}_{n'n}(x) &= 2 (C_{n'} C_n + S_{n'}^- S_n^-) \label{eq:Km} , 
\end{align}
\end{subequations}
where an additional dependence of $K$ on $w_1$, $w_2$, and $w$ is suppressed, with
\begin{subequations}\label{eq:CnSn}
\begin{align}
C_n(x) &= u_n(w_{1}) \cos m \theta_1^\star + u_n(w_{2}) \cos m \theta_2^\star \nonumber \\
& \ \ - u_n(w_{1'}) \cos m \theta_{1'}^\star - u_n(w_{2'}) \cos m \theta_{2'}^\star , \\
S_n^\pm(x) &= u_n(w_{1}) \sin m \theta_1^\star \pm u_n(w_{2}) \sin m \theta_2^\star \nonumber \\
& \ \ - u_n(w_{1'}) \sin m \theta_{1'}^\star \mp u_n(w_{2'}) \sin m \theta_{2'}^\star .
\end{align}
\end{subequations}
All angles in these expressions are chosen such that $\sin \theta_\alpha^\star \geq 0$ (which does not imply a particular sign for $\sin m \theta_\alpha^\star$). 

\subsection{Three-region decomposition}\label{sec:matching}

The form of the collision integral~\eqref{eq:collisiondefinition} does not assume any particular temperature regime. We will now specialize to the low-temperature regime where $\beta\mu \to \infty$. The singularity structure of the collision integral~\eqref{eq:matrixelementsimplified} is encoded in the factor
\begin{align}
\frac{dx}{x |\sin \theta_1^\star \sin \theta_2^\star \, |} \frac{v_F^2}{v_1 v_2} &= \frac{ds}{2 \sqrt{(s-l_1) (s-l_2) (r_1-s) (r_2-s)}} , \label{eq:measure}
\end{align}
which we rewrite on the right-hand side by transforming to a new integration variable $s = x^2$ and resolving the roots of $\sin \theta_1^\star$ and $\sin \theta_2^\star$ near $s\approx 0$
\begin{subequations}\label{eq:lowerboundary}
\begin{align}
l_1 &= \Bigl(\frac{w}{4} \frac{T}{T_F}\Bigr)^2 \biggl[1 - \Bigl(w_1 - \frac{w}{2}\Bigr) \frac{T}{T_F}\biggr] + {\it O}\Bigl(\Bigl(\frac{T}{T_F}\Bigr)^4\Bigr) , \\
l_2 &= \Bigl(\frac{w}{4} \frac{T}{T_F}\Bigr)^2 \biggl[1 - \Bigl(w_2 + \frac{w}{2}\Bigr) \frac{T}{T_F}\biggr] + {\it O}\Bigl(\Bigl(\frac{T}{T_F}\Bigr)^4\Bigr) ,
\end{align}
\end{subequations}
and near $s \approx 1$ 
\begin{subequations}\label{eq:upperboundary}
\begin{align}
r_1 &= 1 + \Bigl(w_1 - \frac{w}{2}\Bigr) \frac{T}{T_F} + {\it O}\Bigl(\Bigl(\frac{T}{T_F}\Bigr)^2\Bigr) , \\
r_2 &= 1 + \Bigl(w_2 + \frac{w}{2}\Bigr) \frac{T}{T_F} + {\it O}\Bigl(\Bigl(\frac{T}{T_F}\Bigr)^2\Bigr) .
\end{align}
\end{subequations}
These roots follow from the condition $\cos^2 \theta_1^\star = 1$ and $\cos^2 \theta_2^\star = 1$ in Eqs.~\eqref{eq:costhetat1} and~\eqref{eq:costhetat2} by setting $\mu=E_F$ and substituting the low-temperature form $k_\alpha = \sqrt{2m^* (w_\alpha T + \mu)/\hbar^2} = k_F \sqrt{1 + w_\alpha T/T_F}$. Since we require $\cos^2 \theta_1^\star \leq 1$ and $\cos^2 \theta_2^\star \leq 1$, the $s$-integration runs between $l_{\rm max} \leq s \leq r_{\rm min}$ with 
\begin{align}
l_{\rm max} &= {\rm max} (l_1, l_2) \\
r_{\rm min} &= {\rm min} (r_1, r_2) .
\end{align}
The lower integral boundary is set by the roots $l_1,l_2$ which are ${\it O}((T/T_F)^2)$ with an absolute splitting  ${\it O}((T/T_F)^3)$, and the upper integral boundary is set by $r_1$ and $r_2$, which are ${\it O}(1)$ with a splitting ${\it O}(T/T_F)$.

We identify three integration regions---an inner region with $s={\it O}((T/T_F)^2)$, an intermediate region with $s={\it O}(1)$, and an outer region with $1-s={\it O}(T/T_F)$, cf.~Fig.~\ref{fig:3}---in which the integrand is expanded in temperature and integrated separately. The overall result is then determined by asymptotic matching of the integral in the overlap regions, which are defined by matching scales $(T/T_F)^2 \ll \delta \ll 1$ and $T/T_F \ll \delta' \ll 1$. Beyond this hierarchy, the scales $\delta$ and $\delta'$ are arbitrary and must cancel from the final matched result. 

The three separate integrals that follow from Eq.~\eqref{eq:matrixelementsimplified} are
\begin{widetext}
\begin{subequations}\label{eq:matching}
\begin{align}
I_{n'n}^{\rm inner}(w_1,w_2,w) &= \frac{1}{32\pi} \int_{l_{\rm max}}^{\delta} \dfrac{ds}{\sqrt{(s-l_1) (s-l_2)}} \Bigl[ 
\underbrace{W_0 \, K_{n'n}^{(m)+}(s)}_{{\rm forward}}
+
\underbrace{W_{\rm ex}(\tilde{s}_>) \, K_{n'n}^{(m)-}(s)}_{{\rm direct}}
\Bigr] \label{eq:Iinner} \\
I_{n'n}^{\rm mid}(w_1,w_2,w) &= \frac{1}{32\pi} \int_{\delta}^{1-\delta'} \dfrac{ds}{s (1-s)} \Bigl[ 
\underbrace{W_{\rm ex}(s) \, K_{n'n}^{(m)+}(s)}_{{\rm exchange}}
+
\underbrace{W_{\rm ho}(s) \, K_{n'n}^{(m)-}(s)}_{{\rm head\text{-}on}}
\Bigr] \label{eq:Imid} \\
I_{n'n}^{\rm outer}(w_1,w_2,w) &= \frac{W_1}{32\pi} \int_{1-\delta'}^{r_{\rm min}} \dfrac{ds}{\sqrt{(r_1-s) (r_2-s)}} \Bigl[ 
\underbrace{K_{n'n}^{(m)+}(s)
+ 
K_{n'n}^{(m)-}(s)}_{{\rm head\text{-}on \ exchange}}
\Bigr] \label{eq:Iouter} .
\end{align}
\end{subequations}
\end{widetext}
The underbraces indicate the kinematic regions that contribute to the integration [cf.~Fig.~\ref{fig:3}]. In writing this equation, we introduce the endpoint values
\begin{align}
    W_0 &= W(0,0) \\
    W_1 &= W(0,1) = W(1,0) \label{eq:W1def}
\end{align}
and define the interaction functions along the different branches as
\begin{align}
    W_{\rm ex}(s) &= W(\sqrt{s},0) \\ 
    W_{\rm ho}(s) &= W(\sqrt{s},\sqrt{1-s}) ,
\end{align}
which have the endpoint limits $W_{\rm ex}(0) = W_0$, $W_{\rm ho}(0) = W_1$ and $W_{\rm ex}(1) = W_{\rm ho}(1) = W_1$. By the duality, we do not need to introduce a separate potential for the direct branch. We discuss the scattering configurations in these regimes in more detail in App.~\ref{app:scatteringconfigurations}. The integration over the direct, exchange, and head-on branches  includes endpoint singularities in the respective forward and head-on direct/exchange overlap regions. The asymptotic kernels are obtained by expanding the integration kernels to leading order in temperature, and we list their explicit forms in App.~\ref{app:asymptotickernels}.

In the integrals in Eq.~\eqref{eq:matching}, the forward  [Eq.~\eqref{eq:kernelforward}] and head-on exchange contribution [Eq.~\eqref{eq:kernelheadonexchange}] do not involve an $s$-dependent kernel and are directly evaluated using the integrals
\begin{align}
\int_{l_{\rm max}}^{\delta} \dfrac{ds}{\sqrt{(s-l_1) (s-l_2)}} 
&= 
\ln \frac{4}{|l_2 - l_1|} + \ln \delta \nonumber \\
&= 2 \ln \frac{4 T_F}{|w| T} + \ln \frac{4 T_F}{|\tilde{w}| T} + \ln \delta , \label{eq:logintegral1} \\
\int_{1-\delta'}^{r_{\rm min}} \dfrac{ds}{\sqrt{(r_1-s) (r_2-s)}} &= \ln \frac{4 \delta'}{|r_1-r_2|} = \ln \frac{4 T_F}{|\tilde{w}| T} + \ln \delta' ,
\end{align}
where we used
\begin{align}
|l_1 - l_2| &
= \Bigl(\frac{w}{4} \frac{T}{T_F}\Bigr)^2 |\tilde{w}| \, \frac{T}{T_F} , \\
|r_1-r_2| &= 
|\tilde{w}| \frac{T}{T_F} .
\end{align}
The logarithmic temperature divergence in Eq.~\eqref{eq:logintegral1} stems from the fact that while the roots $l_1$ and $l_2$ are ${\it O}(T^2/T_F^2)$, their separation is ${\it O}(T^3/T_F^3)$. 

The remaining direct, exchange, and head-on integrals have logarithmic endpoint singularities that must be extracted. For the exchange and head-on branches, this separation is done schematically using
\begin{align}
&\int_{\delta}^{1-\delta'} ds \, \frac{G(s)}{s (1-s)} 
= - G(0) \ln \delta - G(1) \ln \delta' \nonumber \\
&\qquad + \int_{0}^{1} ds \, \biggl[\frac{G(s)}{s (1-s)} - \frac{G(0)}{s} - \frac{G(1)}{1-s}\biggr] , \label{eq:sintegrationschema}
\end{align}
where $G(s)$ is the numerator, and in the second line we extend the integration over the full interval $[0,1]$ for the finite remainder. 

The direct-branch integration is brought to the same form by transforming to a new integration parameter $\tilde{s} = 1 - (wT/4 T_F)^2/s$. This gives, approximating $\sqrt{(s-l_1) (s-l_2)} \approx s - (wT/4T_F)^2$,
\begin{align}
\int_{l_{\rm max}}^{\delta} \dfrac{ds}{\sqrt{(s-l_1) (s-l_2)}} \,G\biggl(\frac{w T}{4 T_F \sqrt{s}} \biggr) 
&= \int_{\tilde{s}_-}^{\tilde{s}_+} d\tilde{s} \dfrac{G(\sqrt{1-\tilde{s}})}{\tilde{s} (1-\tilde{s})} ,
\end{align}
where the integral boundaries follow from solving the full Eq.~\eqref{eq:constraint} for $\tilde{s}$, which gives to linear order $\tilde{s}(l_{1/2}) = \mp \tilde{w} T/(4 T_F)$ and hence $\tilde{s}_- = \max\{ \tilde{s}(l_{1}), \tilde{s}(l_{2}) \} = |\tilde{w}| T/(4 T_F)$, as well as $\tilde{s}_+ = \tilde{s}(\delta) = 1 - w^2 T^2/(16 T_F^2 \delta)$, which is close to $1$ since $T^2/T_F^2 \ll \delta$. 

Combining these results, we have
\begin{align}\label{eq:integrandresult}
&I_{n'n}(w_1,w_2,w) = 
 \underbrace{\dfrac{W_0}{8\pi} \Delta_{n'} \Delta_n \Biggl[ \ln \dfrac{4 T_F}{|w| T} + \ln \dfrac{4 T_F}{|\tilde{w}| T} \Biggr]}_{{\rm forward}} \nonumber \\
 &+ \dfrac{W_1}{8\pi} \Biggl[ 
\underbrace{ \begin{Bmatrix}
\Delta_{n'} \Delta_n \\
D_{n'} D_n 
\end{Bmatrix}
\ln \dfrac{4 T_F}{|w| T}}_{{\rm head\text{-}on \ direct}} + 
\underbrace{ \begin{Bmatrix}
\Delta_{n'} \Delta_n \\
E_{n'} E_n 
\end{Bmatrix}
\ln \dfrac{4 T_F}{|\tilde{w}| T}}_{{\rm head\text{-}on \ exchange}} \Biggr] 
 \nonumber \\
 &+ {\cal R}_{n'n}^{(m)} ,
\end{align}
where we define
\begin{subequations}\label{eq:defDENSigma}
\begin{align}
\Delta_n &= u_n(w_1) + u_n(w_2) - u_n(w_{1'}) - u_n(w_{2'}) , \\
D_n &= u_n(w_1) - u_n(w_2) - u_n(w_{1'}) + u_n(w_{2'}) , \\
E_n &= u_n(w_1) - u_n(w_2) + u_n(w_{1'}) - u_n(w_{2'}) ,\\
\Sigma_n &= u_n(w_1) + u_n(w_2) + u_n(w_{1'}) + u_n(w_{2'}) .
\end{align}
\end{subequations}
In the curly brackets in Eq.~\eqref{eq:integrandresult}, the upper (lower) entry  holds for even $m$ (odd $m$), and we indicate the origin of the different logarithmic endpoint singularities in underbraces. Particle number and energy conservation imply $\Delta_0 = \Delta_1 = 0$. In addition, $D_0 = E_0 = 0$ because $u_0=1$ is constant, while $\Sigma_0=4$. Equation~\eqref{eq:integrandresult} shows that the leading logarithmic temperature scaling arises from the endpoint regions with forward and head-on direct and exchange scattering. The nonlogarithmic remainder in Eq.~\eqref{eq:integrandresult} is written concisely as
\begin{widetext}
\begin{align}\label{eq:Rnpn}
{\cal R}_{n'n}^{(m)} &=
\Omega_\Delta^{(m)}[W] \, \frac{\Delta_{n'} \Delta_n}{4 \pi} + \Omega_{DE}^{(m)}[W] \, \frac{D_{n'} D_n + E_{n'} E_n}{8\pi} 
+ \Omega_\Sigma^{(m)}[W] \, \frac{\Sigma_{n'} \Sigma_n}{4 \pi}
\end{align}
with
\begin{subequations}\label{eq:residualcoeffs}
\begin{align}
\Omega_\Delta^{(m)}[W] &= 
\begin{cases}
\displaystyle - \frac{1}{4} \int_{0}^{1} \frac{ds}{s} \, U_{m-1}^2\bigl(\sqrt{s}\bigr) \bigl[ 2 W_{\rm ex}(s) + W_{\rm ho}(s)\bigr] \\[2ex]
\qquad \displaystyle + \dfrac{1}{2} \int_0^1 \frac{ds}{s} \, \bigl[(W_{\rm tot}(s) - 2 W_1) + (W_{\rm ex}(s) - W_0)
\bigr]
& m \ {\rm even} \\[2ex]
\displaystyle \frac{1}{2} \int_{0}^{1} \frac{ds}{s} \, U_{m-1}^2\bigl(\sqrt{s}\bigr) 
\bigl[W_{\rm ex}(s) - W_0 \bigr] 
& m \ {\rm odd}
\end{cases} 
\\[2ex]
\Omega_{DE}^{(m)}[W] &= \begin{cases}
\displaystyle \frac{1}{2} \int_{0}^{1} \frac{ds}{s} \, U_{m-1}^2\bigl(\sqrt{s}\bigr) 
\, W_{\rm ex}(s)
& m \ {\rm even} \\[2ex]
\displaystyle \frac{1}{2} \int_{0}^{1} \frac{ds}{s} \, U_{m-1}^2\bigl(\sqrt{s}\bigr)\bigl[W_{\rm tot}(s) 
- 2 W_1\bigr] 
& m \ {\rm odd}
\end{cases} \label{eq:defPhiDE} \\
\Omega_\Sigma^{(m)}[W] &= \begin{cases}
\displaystyle \frac{1}{4} \int_{0}^{1} \frac{ds}{s} \, U_{m-1}^2\bigl(\sqrt{s}\bigr)
\, W_{\rm ho}(s)
& m \ {\rm even} \\[2ex]
0 & m \ {\rm odd} ,
\end{cases}
\end{align}
\end{subequations}
\end{widetext}
where we define a total effective potential
\begin{align}
 W_{\rm tot}(s) &= W_{\rm ex}(1-s) + W_{\rm ho}(s) , \label{eq:Wtotdef}
\end{align}
which includes contributions from the head-on and the direct-exchange branch and has endpoint limits $W_{\rm tot}(0) = 2 W_1$ and $W_{\rm tot}(1) = W_0+W_1$. In deriving these expressions from the general form~\eqref{eq:sintegrationschema}, we use the Chebyshev identity $1-T_m^2(\sqrt{s}) = (1-s) U_{m-1}^2(\sqrt{s})$ as well as two identities valid for odd $m$, $T_m^2(\sqrt{1-s}) = 1-T_m^2(\sqrt{s})$ and 
\begin{align}
    \int_0^1 \frac{ds}{s} [U_{m-1}^2(\sqrt{s}) - 1] &= 0 .
\end{align}
The integrals~\eqref{eq:residualcoeffs} contain all information on the interaction potential. For further reference, we note the large-$m$ limit of the integral~\eqref{eq:defPhiDE} for odd $m$, 
\begin{align}
 \Omega_{DE}^{(m)}[W] \simeq \frac{1}{2} (W_0-W_1) \ln m \label{eq:Omega11asymptotic} ,
\end{align}
which has an amplitude set by the endpoint values $W_0$ and $W_1$. This limit is derived by substituting $s = \sin^2 y$ in Eq.~\eqref{eq:residualcoeffs} and using the asymptotic form of the Chebyshev polynomials in the integrand.

\subsection{Long-lived odd matrix elements}\label{sec:matchingderivation}

For even $m$, since $\Delta_0=\Delta_1=0$, by Eq.~\eqref{eq:integrandresult} only elements with $n',n \geq 2$  have a logarithmic scaling correction. By Eq.~\eqref{eq:Rnpn}, for $n=0$, finite corrections are proportional to $\Sigma_{n'} \Sigma_n$. This term sets the $T^2$ Fermi-liquid relaxation rate of the lowest radial elements, and it originates from the head-on scattering branch, as discussed in the introduction.

Likewise, for odd $m$, by Eq.~\eqref{eq:integrandresult} only elements with $n',n \geq 1$ will have a logarithmic divergence at this order of the analysis. By Eq.~\eqref{eq:residualcoeffs} odd coefficients with $n'=0$ or $n=0$ also do not have a finite remainder since \mbox{$\Omega_\Sigma^{(m)}=0$}. For $n'=0$, we obtain to leading nonzero order
\begin{align}\label{eq:integrandresultodd}
&I_{00}(w_1,w_2,w) = {\cal R}_{00} \nonumber \\
&+
\frac{c_m^2 W_1}{2} \frac{T^2}{T_F^2}
\Biggl[
\underbrace{
\frac{D_1^2}{4\pi} \ln \frac{4T_F}{|w|T}
}_{{\rm head\text{-}on\ direct}}
+
\underbrace{
\frac{E_1^2}{4\pi} \ln \frac{4T_F}{|\tilde{w}|T}
}_{{\rm head\text{-}on\ exchange}}
\Biggr] ,
\\
&I_{0n}(w_1,w_2,w) = {\cal R}_{0n} \nonumber \\
&
- \frac{c_m W_1}{2} \frac{T}{T_F}
\Biggl[
\underbrace{
\frac{D_1 D_n}{4\pi} \ln \frac{4T_F}{|w|T}
}_{{\rm head\text{-}on\ direct}}
+
\underbrace{
\frac{E_1 E_n}{4\pi} \ln \frac{4T_F}{|\tilde{w}|T}
}_{{\rm head\text{-}on\ exchange}}
\Biggr] ,
\end{align}
where we used $w = \pi D_1/(2\sqrt{3})$ and $\tilde{w} = \pi E_1/(2\sqrt{3})$ and define
\begin{align}
 c_m &= \frac{m^2 \pi}{2 \sqrt{3}} . \label{eq:defcm}
\end{align}
In contrast to the generic matrix elements in Eq.~\eqref{eq:integrandresult}, there is no forward-scattering contribution at this order. The leading logarithmic terms arise entirely from the head-on direct and head-on exchange endpoint regions, as indicated.

Using the change of variables $s\leftrightarrow 1-s$ to combine the direct and exchange contributions as well as the two endpoints of the head-on contribution, the finite remainders are
\begin{widetext}
\begin{align}
{\cal R}_{00} &= \frac{T^2}{T_F^2} \Omega_{00}^{(m)}[W] \, \frac{D_1^2 + E_1^2}{8\pi} , \\
{\cal R}_{0n} &= \frac{T}{T_F} \Omega_{01}^{(m)}[W] \, \frac{D_1 D_n + E_1 E_n}{8\pi}
\end{align}
with
\begin{align}
&\Omega_{00}^{(m)}[W]
=
\frac{\pi^2 m^2}{24} 
\int_0^1 ds\,
\dfrac{T_m^2(\sqrt{s})}{s^2} \, \bigl[W_{\rm tot}(s)-2W_1\bigr]
+
\frac{\pi^2 m^2}{24} 2 W_1 
\int_0^1 ds\,
\biggl[
\dfrac{T_m^2(\sqrt{s})}{s^2}
-
\dfrac{m^2}{s} 
\biggr]
,
\label{eq:R00odd}
\\
&\Omega_{01}^{(m)}[W] 
=
- \frac{\pi m}{4\sqrt{3}} 
\int_0^1 ds\,
\dfrac{
T_m(\sqrt{s})U_{m-1}(\sqrt{s})
}{s^{3/2}}
\bigl[
W_{\rm tot}(s) - 2 W_1
\bigr]
- \frac{\pi m}{4\sqrt{3}} 2 W_1
\int_0^1 ds\,
\biggl[
\dfrac{
T_m(\sqrt{s})U_{m-1}(\sqrt{s})
}{s^{3/2}}
-
\dfrac{m}{s} 
\biggr] . 
\label{eq:R0nodd}
\end{align}
\end{widetext}
For further reference, at large $m$, these integrals scale as 
\begin{align}
\Omega_{00}^{(m)}[W] &\simeq - 2 c_m^2 W_1 \ln m, \label{eq:Omega00asymptotic} \\
\Omega_{01}^{(m)}[W] &\simeq 2 c_m W_1 \ln m \label{eq:Omega01asymptotic} ,
\end{align}
respectively.

\subsection{Summary of the low-temperature collision matrix}\label{sec:asymptoticresults}

We now collect the low-temperature form of the collision matrix elements. The coefficients take a simple product form that factorizes into angular and radial parts. The radial coefficients depend on the indices $n'$ and $n$ but neither on the angular harmonic index $m$ nor the form of the interaction potential, and the angular coefficients are averages over the interaction potential with a kinematic factor that depends on $m$. We define the matrices over the radial basis functions as
\begin{subequations}\label{eq:resFHln}
\begin{align}
\bar{\Phi}_{n'n}^{DE} &= \frac{1}{4\pi} \Bigl\langle D_{n'} D_n \ln \frac{4}{|w|} \Bigr\rangle_F = \frac{1}{4\pi} \Bigl\langle E_{n'} E_n \ln \frac{4}{|\tilde{w}|} \Bigr\rangle_F , \label{eq:defHbar} \\
\bar{\Phi}_{n'n}^\Delta &= \frac{1}{4\pi} \Bigl\langle \Delta_{n'} \Delta_n \ln \frac{4}{|w|} \Bigr\rangle_F = \frac{1}{4\pi} \Bigl\langle \Delta_{n'} \Delta_n \ln \frac{4}{|\tilde{w}|} \Bigr\rangle_F , \label{eq:defFbar}
\end{align}
\end{subequations}
where the Fermi average is defined in Eq.~\eqref{eq:Fermiintegral}. Because of the parity of the radial basis functions, the matrix coefficients are nonzero only if $n+n'$ is even. Their evaluation is discussed in App.~\ref{app:logfermiintegrals}. Likewise, we define the Fermi averages without the logarithm, 
\begin{subequations}
\begin{alignat}{4} 
\Phi_{n'n}^{DE} &= \frac{1}{4\pi} \langle D_{n'} D_n \rangle_F = \frac{1}{4\pi} \langle E_{n'} E_n \rangle_F , \\
\Phi_{n'n}^\Delta &= \frac{1}{4\pi} \langle \Delta_{n'} \Delta_n \rangle_F , \\
\Phi_{n'n}^\Sigma &= \frac{1}{4\pi} \langle \Sigma_{n'} \Sigma_n \rangle_F .
\end{alignat}
\end{subequations}
These coefficient matrices are computed in App.~\ref{app:Fermiintegrals}. They are band-diagonal matrices with entries for \mbox{$n' = n, n\pm 2$}. By contrast, because of the factors $\ln (4/|w|)$ and $\ln (4/|\tilde{w}|)$, $\bar{\Phi}_{n'n}^{DE}$ and $\bar{\Phi}_{n'n}^{\Delta}$ are not band diagonal.

For odd $m$, we first list the asymptotic form for the entries of the first row and column
\begin{subequations}\label{eq:L00L0n}
\begin{align}
L_{00}^{(m)} 
&= \frac{T^4}{\hbar T_F^3} \Bigl[a_{00} \Lambda_T + b_{00} \Bigr] ,
\\
L_{0n}^{(m)} 
&= \frac{T^3}{\hbar T_F^2} \Bigl[a_{n} \Lambda_T + b_{n} \Bigr] 
\end{align}
\end{subequations}
with
\begin{subequations}\label{eq:a}
\begin{align}
a_{00} &= c_m^2 W_1 \Phi_{11}^{DE} 
\label{eq:a00} \\
a_n &= - c_m W_1 \Phi^{DE}_{1n} 
\label{eq:a0n} 
\end{align}
\end{subequations}
and
\begin{subequations}\label{eq:b}
\begin{align}
b_{00} &= \Omega_{00}^{(m)}[W] \Phi_{11}^{DE} + c_m^2 W_1 \bar{\Phi}_{11}^{DE} \label{eq:b00} \\
b_{n} &= \Omega_{01}^{(m)}[W] \Phi^{DE}_{1n} - c_m W_1 \bar{\Phi}^{DE}_{1n} \label{eq:b0n} ,
\end{align}
\end{subequations}
where $\Omega_{00}^{(m)}$ and $\Omega_{01}^{(m)}$ are defined in Eqs.~\eqref{eq:R00odd} and~\eqref{eq:R0nodd}, respectively. 

For all remaining matrix elements---i.e., for all odd $m$ with $n,n'\geq 1$ and all even $m$---we obtain instead the result (the even-$m$ element $L_{01}^{(m)}$ vanishes at this order and its leading ${\it O}(T^3)$ term is derived in App.~\ref{app:L01})
\begin{align}
L_{n'n}^{(m)} &= \frac{T^2}{\hbar T_F} \Bigl[A_{n'n} \Lambda_T + B_{n'n} \Bigr] . \label{eq:collisionmatrix}
\end{align}
The coefficients of this expression are set by the Fermi averages
\begin{align}
A_{n'n} &= \begin{cases}
(W_0 + W_1) \Phi_{n'n}^\Delta & m \ {\rm even} \\
W_0 \Phi_{n'n}^\Delta + W_1 \Phi_{n'n}^{DE} & m \ {\rm odd} 
\end{cases} \label{eq:defA}
\end{align}
and
\begin{align}
B_{n'n} &= \begin{cases}
(W_0 + W_1) \bar{\Phi}_{n'n}^\Delta + \Omega_\Delta^{(m)}[W] \Phi_{n'n}^\Delta \\[1ex]
\quad + \Omega_{DE}^{(m)}[W] \Phi_{n'n}^{DE} + \Omega_\Sigma^{(m)}[W] \Phi_{n'n}^\Sigma & m \ {\rm even} \\[2ex]
W_0 \bar{\Phi}_{n'n}^\Delta + W_1 \bar{\Phi}_{n'n}^{DE} \\[1ex]
\quad + \Omega_\Delta^{(m)}[W] \Phi_{n'n}^\Delta + \Omega_{DE}^{(m)}[W] \Phi_{n'n}^{DE} & m \ {\rm odd} ,
\end{cases} \label{eq:defB}
\end{align}
where expressions for the angular coefficients $\Omega_{\Delta,DE,\Sigma}^{(m)}$ are listed in Eq.~\eqref{eq:residualcoeffs}.

\section{Long-lived odd eigenmode}\label{sec:eigenmodes}

In this section, we derive the relaxation rates of long-lived odd modes, including Eq.~\eqref{eq:oddmoderesult} in the introduction. The structure of the odd-mode collision matrix~\eqref{eq:structure} shows directly that for any interaction potential the spectrum splits into a single slow eigenvalue of ${\it O}(T^4)$ and an infinite set of fast eigenvalues of ${\it O}(T^2)$. The slow eigenmode follows from the Schur complement of the lower block $n,n'\geq 1$, which is ${\it O}(T^4)$, while the remaining modes follow from diagonalizing the lower block $A \Lambda_T + B$, which is ${\it O}(T^2)$~\cite{matrixcookbook}. The structure of the eigenmode already becomes clear by including the first two radial modes. We discuss this structure first in Sec.~\ref{sec:twobasis}, which is followed by a full discussion including all radial modes in Sec.~\ref{sec:eigenmodelgeneral}, which establishes rapid convergence in the number of radial modes.

\subsection{Two-moment variational solution}\label{sec:twobasis}

Consider the first $2\times 2$ block of the collision matrix~\eqref{eq:structure}
\begin{align}
 L^{(m)} &= \frac{T^2}{\hbar T_F} 
 \begin{pmatrix}
 \frac{T^2}{T_F^2} \bigl(a_{00} \Lambda_T + b_{00}\bigr) & \frac{T}{T_F} \bigl(a_1 \Lambda_T + b_1\bigr) \\[1ex]
 \frac{T}{T_F} \bigl(a_1 \Lambda_T + b_1\bigr) & A_{11} \Lambda_T + B_{11}
 \end{pmatrix} , 
\end{align}
where the coefficients $a_{00}$, $b_{00}$, $a_1$, and $b_1$ are defined in Eqs.~\eqref{eq:a} and~\eqref{eq:b}. We note the identities
\begin{subequations}
\label{eq:logcancellationcoeff}
\begin{align}
 a_{00} &= c_m^2 A_{11} \\
 a_n &= - c_m A_{1n} ,
\end{align}
\end{subequations}
which follow from energy conservation: Since $\Delta_{1} = 0$, one has $\Phi^{\Delta}_{1n}=0$ for all $n$, which ensures $A_{1n} = W_1 \Phi^{DE}_{1n}$ [Eq.~\eqref{eq:defA}], and Eq.~\eqref{eq:logcancellationcoeff} follows directly from Eqs.~\eqref{eq:a00} and~\eqref{eq:a0n}. 

The deformation in Eq.~\eqref{eq:odddeformation} reads in vector representation 
\begin{align}
{\bf v} &= \begin{pmatrix} 1 \\ c_m \xi \frac{T}{T_F} \end{pmatrix} ,
\end{align}
where we use that $c_m u_1(w) = m^2 w/2$. The variational estimate for the relaxation rate of this state is
\begin{align}
\gamma_m(\xi) &= {\bf v}^T L^{(m)} {\bf v} = \frac{T^4}{\hbar T_F^3} \Bigl[c_m^2 (A_{11} \Lambda_T + B_{11}) \xi^2 \nonumber \\
&\quad + 2 c_m (a_1 \Lambda_T + b_1) \xi + a_{00} \Lambda_T + b_{00}\Bigr] , \label{eq:variation}
\end{align}
where we do not include the norm of ${\bf v}$ since it varies only at higher order in temperature, ${\bf v}^T {\bf v} = 1 + {\it O}(T^2/T_F^2)$. Substituting the coefficients in Eq.~\eqref{eq:variation} in terms of the angle integrals $\Omega_{00}^{(m)}$, $\Omega_{01}^{(m)}$, and $\Omega_{DE}^{(m)}$ gives Eq.~\eqref{eq:oddmoderesult}. Here, the logarithmic coefficients combine directly $(c_m^2 A_{11} \xi^2 + 2 c_m a_1 \xi + a_{00}) \Lambda_T = c_m^2 A_{11} (\xi-1)^2 \Lambda_T = c_m^2 \Phi_{11}^{DE} (\xi-1)^2 W_1 \Lambda_T$. The term in the finite coefficients involving $\bar{\Phi}_{11}^{DE}$ combines in the same way, giving $c_m^2 \Phi_{11}^{DE} (\xi-1)^2 W_1 h$, with $h=\bar{\Phi}_{11}^{DE}/\Phi_{11}^{DE}$. The remaining terms evaluate to
\begin{align}
&c_m^2 \xi^2 \Omega_{DE}^{(m)} + 
 2 c_m \xi \Omega_{01}^{(m)} + \Omega_{00}^{(m)} \nonumber \\
&= \frac{c_m^2}{2} \int_0^1 \frac{ds}{s} \biggl[W_{\rm tot}(s) U_{m-1}^2(\sqrt{s}) \bigl(\xi - \bar{\xi}(s)\bigr)^2 \nonumber \\
&\qquad - 2 W_1 \bigl(\xi - 1\bigr)^2\biggr] ,
\end{align}
which, combined with $c_m^2 \Phi_{11}^{DE}/2 = \pi^3 m^4/15$, gives Eq.~\eqref{eq:oddmoderesult}.

Minimizing Eq.~\eqref{eq:variation} [or, equivalently, Eq.~\eqref{eq:oddmoderesult}] gives
\begin{align}
\xi_m &= - \frac{a_1 \Lambda_T + b_1}{c_m (A_{11} \Lambda_T + B_{11})} 
= 1 - \frac{2 {\cal K}_m}{{\cal S}_m} , \label{eq:xistar}
\end{align}
where we introduce the shorthands
\begin{align}
{\cal K}_m &= \frac{b_1 + c_m B_{11}}{c_m \Phi_{11}^{DE}} 
\nonumber \\
&= \int_0^1 ds \, W_{\rm tot}(s) U_{m-1}(\sqrt{s}) \Bigl[\frac{d}{ds} \frac{T_{m}(\sqrt{s})}{m\sqrt{s}}\Bigr] \label{eq:Kmcoeff}
\end{align}
and
\begin{align}
{\cal S}_m &= 2 \frac{A_{11} \Lambda_T + B_{11}}{\Phi_{11}^{DE}}  \nonumber \\
& = 2 W_1 \bar{\Lambda}_T + \int_0^1 \frac{ds}{s} \, U_{m-1}^2(\sqrt{s}) \bigl[ W_{\rm tot}(s) - 2 W_1\bigr] . \label{eq:Smcoeff}
\end{align}
For asymptotically small temperatures $\xi_m \to 1$, and the second term in Eq.~\eqref{eq:xistar} describes inverse-logarithmic corrections to this result, which are formally suppressed by a factor of ${\cal K}_m/[W_1\Lambda_T]$. From the scaling of the coefficient $b_1 \sim \Omega_{01}^{(m)}$ [cf.~Eq.~\eqref{eq:Omega01asymptotic}], the asymptotic form of ${\cal K}_m$ for large $m$ is ${\it O}(\ln m)$. Hence, the size of the subleading inverse logarithmic corrections compared to the leading term is governed by $\ln m/\Lambda_T$. As discussed in the introduction, within the range of applicability of the low-temperature expansion $m \lesssim \sqrt{T_F/T}$, this correction can be as large as $\ln m/\Lambda_T = 1/2$. The long-lived odd modes carry the largest relative dressing at small $m$ and then cross over with increasing $m$ to a rigid Fermi-surface deformation with $\xi_m \simeq 0$. A special case is $m=1$, for which the derivative in the integrand of Eq.~\eqref{eq:Kmcoeff} is zero, and ${\cal K}_{m=1}$ vanishes for all interaction potentials. This implies $\xi_{m=1} = 1$, as required by momentum conservation.

With the expression~\eqref{eq:xistar}, Eq.~\eqref{eq:variation} or Eq.~\eqref{eq:oddmoderesult} takes the form
\begin{align}
&\gamma_m(\xi) \nonumber \\* 
&= 2 c_m^2 \Phi_{11}^{DE} \frac{T^4}{\hbar T_F^3} \, \biggl[{\cal I}_m - \frac{1}{2} (1 - \xi_m) {\cal K}_m + \frac{{\cal S}_m}{4} \bigl(\xi - \xi_m\bigr)^2 \biggr] \label{eq:oddmoderelax}
\end{align}
with
\begin{align}
{\cal I}_m &= \frac{b_{00} + 2 c_m b_1 + c_m^2 B_{11}}{2 c_m^2 \Phi_{11}^{DE}} \nonumber \\
&= \int_0^1 ds \, s \, W_{\rm tot}(s) \Bigl[\frac{d}{ds} \frac{T_{m}(\sqrt{s})}{m\sqrt{s}}\Bigr]^2 . \label{eq:Imcoeff}
\end{align}

\subsection{General expression}\label{sec:eigenmodelgeneral}

The general expression for the eigenvalue of the slow mode beyond the $n=0,1$ variational treatment follows from the Schur complement of the lower block $n,n'\geq 1$, which is
\begin{align}
\gamma_m &= \frac{T^4}{\hbar T_F^3} \Bigl[a_{00} \Lambda_T + b_{00} \nonumber \\
&\qquad\qquad- \bigl({\bf a} \Lambda_T + {\bf b}\bigr)^T \Bigl(A \Lambda_T + B\Bigr)^{-1} \bigl({\bf a} \Lambda_T + {\bf b}\bigr)\Bigr] \nonumber \\
&= \frac{T^4}{\hbar T_F^3} \Bigl[
\bigl(a_{00} - c_m^2 A_{11} \bigr) \Lambda_T + 2 c_m^2 {\cal I}_m \Phi_{11}^{DE} 
\nonumber \\
&\qquad \qquad - {\bf r}^T \bigl(A \Lambda_T + B\bigr)^{-1} {\bf r}\Bigr] . \label{eq:gammaslow}
\end{align}
In the second line, we make use of the identities~\eqref{eq:logcancellationcoeff} to separate the logarithmic terms by defining ${\bf a} \Lambda_T + {\bf b} = - c_m (A \Lambda_T+ B){\bf e}_1 + {\bf r}$ with an auxiliary vector \begin{align}
 {\bf r} &= c_m B {\bf e}_1 + {\bf b} ,
\end{align}
where we write ${\bf e}_1=(1,0,0,\ldots)$, such that $(A {\bf e}_1)_n = A_{n1}$ and $(B {\bf e}_1)_n = B_{n1}$. 

Superficially, the expression~\eqref{eq:gammaslow} has a logarithmic temperature divergence ${\it O}(\Lambda_T)$. However, by Eq.~\eqref{eq:logcancellationcoeff}, its coefficient $a_{00} - c_m^2 A_{11} = a_{00} - {\bf a}^T A^{-1} {\bf a}$ vanishes  identically for any interaction potential. The remaining term on the first line of Eq.~\eqref{eq:gammaslow} describes the leading contribution to the slow-mode relaxation rate as $1/\Lambda_T \to 0$. The correction term in the second line of Eq.~\eqref{eq:gammaslow} involves the inverse radial collision matrix. Note that, unlike for a 3D Fermi liquid, where the matrix $L_{n'n}$ is only nonzero along the diagonal $n=n'$ and the bands $n=n'\pm 2$~\cite{brooker68,repplinger25,taillat25,taillat26,kurkjian26}, the 2D collision matrix is not sparse due to the contribution of the logarithmic energy averages $\bar{\Phi}_{n'n}^{DE}$ and $\bar{\Phi}_{n'n}^\Delta$, cf.~Eqs.~\eqref{eq:defHbar} and~\eqref{eq:defFbar}. The full inverse thus cannot be obtained in closed form, for example, by a continued fraction expansion. Truncated to the first two radial modes, Eq.~\eqref{eq:gammaslow} reproduces Eq.~\eqref{eq:oddmoderelax} evaluated at its minimum value $\xi = \xi_m$, which is seen from ${\bf e}_1^T {\bf r} = c_m \Phi_{11}^{DE} {\cal K}_m$ and the definition of ${\cal S}_m$.

The structure of the odd slow mode follows from the eigenvalue equation $L_{n'n}^{(m)} \psi_{n} = \gamma_m \psi_{n'}$. Setting $\psi = (1,\boldsymbol{\psi}^{(m)})$, we have $(A \Lambda_T + B) {\boldsymbol{\psi}^{(m)}} = - ({\bf a} \Lambda_T + {\bf b}) T/T_F$ at small temperatures, which is solved as
\begin{align}
 \psi_n^{(m)} &= \frac{T}{T_F} \Bigl(c_m \delta_{n1} - \bigl[(A \Lambda_T + B)^{-1} {\bf r}\bigr]_n \Bigr) . \label{eq:deformation}
\end{align}
The first term describes the fully dressed collinearly protected mode, and the second term quantifies how much this protection is relaxed by noncollinear processes (for $n=1$) as well as the admixture of higher radial modes (for $n\geq 2$). Equation~\eqref{eq:deformation} requires $c_m T/T_F \sim m^2 T/T_F \ll 1$ to be a small correction, which provides an additional derivation of the temperature constraint $m \lesssim \sqrt{T_F/T}$ derived from the matched expansion.

Further progress can be made for the first two terms in an expansion of Eq.~\eqref{eq:gammaslow} in inverse powers $1/\Lambda_T$. The contribution of $\bar{\Phi}^{DE}_{1n}$ to the vector ${\bf r}$ cancels, which implies ${\bf r} = (\Omega_{01}^{(m)} + c_m \Omega_{DE}^{(m)}) \Phi^{DE} {\bf e}_1$, and hence
\begin{align}
 A^{-1} {\bf r} &= \frac{1}{W_1} (\Omega_{01}^{(m)} + c_m \Omega_{DE}^{(m)}) \, {\bf e}_1 ,
\end{align}
such that
\begin{align}
&(A \Lambda_T + B)^{-1} {\bf r} = \frac{1}{\Lambda_T} A^{-1} {\bf r} - \frac{1}{\Lambda_T^2} A^{-1} B A^{-1} {\bf r} + \ldots \nonumber \\
&= \frac{\Omega_{01}^{(m)} + c_m \Omega_{DE}^{(m)}}{W_1 \Lambda_T} \Bigl[{\bf e}_1 - \frac{1}{\Lambda_T} A^{-1} B {\bf e}_1 + {\it O}\bigl(1/\Lambda_T^2\bigr)\Bigr] . 
\end{align}
In an expansion in powers of the inverse logarithmic temperature $1/\Lambda_T$, up to ${\it O}(1/\Lambda_T^2)$ the rate includes only the $n=1$ component, which  relaxes the mode from the fully dressed limit $\xi=1$. Only at ${\it O}(1/\Lambda_T^3)$ do higher radial modes $n\geq 2$ contribute. We find such corrections numerically small and the ansatz~\eqref{eq:odddeformation} correct to within a few percent [Fig.~\ref{fig:5}(c)]. Resumming the $n=1$ terms is equivalent to minimizing over $\xi$ in Eq.~\eqref{eq:oddmoderesult}.

\subsection{Large-$m$ limit and comparison with \citet{ledwith19}}\label{sec:projectedrate}

We now compare the results for the odd-mode relaxation rate with the rate obtained by \citet{ledwith19}. Their derivation is based on the $2\times 2$ collision matrix evaluated to leading logarithmic order (cf.~Eqs.~(63) and (95) of Ref.~\cite{ledwith19}). Their calculation of the odd collision matrix corresponds, written here in our notation, to
\begin{align}
 L^{(m)} &= \frac{T^2}{\hbar T_F} 
 \begin{pmatrix}
 \frac{T^2}{T_F^2} a_{00} (\Lambda_T - \ln m^2) & \frac{T}{T_F} a_1 (\Lambda_T - \ln m^2) \\[1ex]
 \frac{T}{T_F} a_1 (\Lambda_T - \ln m^2) & A_{11} \Lambda_T 
 \end{pmatrix} . \label{eq:LmLedwith}
\end{align}
The $00$ and the $01$ matrix elements of Eq.~\eqref{eq:LmLedwith} include a logarithm of the angular harmonic index $m$, which arises from cutting off the integration region that gives rise to the logarithmic temperature term. 

To compare with these results, we expand our matrix elements to leading order at large $\ln m \gg 1$, which gives
\begin{subequations}
\begin{align}
 \frac{\hbar T_F^3}{T^4} L_{00} &= \, \ \ c_m^2 \Phi_{11}^{DE} W_1 \bigl[ \Lambda_T - 2 \ln m \bigr] \\
 \frac{\hbar T_F^2}{T^3} L_{01} &= - c_m \Phi_{11}^{DE} W_1 \bigl[ \Lambda_T - 2 \ln m \bigr] \\
 \frac{\hbar T_F}{T^2} L_{11} &= \ \ \ \ \ \ \Phi_{11}^{DE} W_1  \Bigl[ \Lambda_T + \frac{1}{2} \frac{W_0 - W_1}{W_1} \ln m \Bigr].
\end{align}
\end{subequations}
The last term in each expression follows directly from the asymptotic forms of $\Omega_{00}^{(m)}$, $\Omega_{01}^{(m)}$, and $\Omega_{DE}^{(m)}$ in Eqs.~\eqref{eq:Omega00asymptotic},~\eqref{eq:Omega01asymptotic},~and~\eqref{eq:Omega11asymptotic}, respectively. The first two matrix elements agree with Ref.~\cite{ledwith19}, and the last element agrees up to the term that is proportional to the difference between the collinear scattering amplitude $W_1$ and the forward-scattering amplitude $W_0$, which vanishes only if they are equal (for example, for a constant interaction). Expanding the eigenvalue equation for $\Lambda_T \gg \ln m$, we find that the logarithmic temperature scaling of the odd-mode eigenvalue cancels, but a logarithmic $m$-divergence remains, which gives 
\begin{align}
\gamma_{m\gg1}^{(0)} &= \frac{\pi^3}{15} \frac{T^4}{\hbar T_F^3} (W_0 + 3 W_1) m^4 \ln m + {\it O}(m^4), \label{eq:closedformasymptotic}
\end{align}
where the superscript $(0)$ indicates that the rate is evaluated at full thermal dressing $\xi=1$. The functional form of Eq.~\eqref{eq:closedformasymptotic} agrees with the result by \citet{ledwith19}, but its prefactor $W_0 + 3 W_1$ differs from the coefficient $4W_1$ of Ref.~\cite{ledwith19} (cf.~Eq.~(96) in Ref.~\cite{ledwith19}). Their result is proportional to the collinear head-on amplitude $|V_\star|^2$ [cf.~Eq.~(15) in that work, which gives $W_1 = N_0^2 |V_\star|^2/2$ with the spin average described there], whereas we find an admixture of the collinear head-on amplitude $W_1$ and the forward-scattering amplitude $W_0$. For a constant interaction, these results agree, shown as a green dotted line in Fig.~\ref{fig:1}.

The reduction in the dressing that increases the lifetime of the odd mode for larger $m$ compared to full dressing can also be discussed starting from the asymptotic expansion. To this end, consider the asymptotic form of Eq.~\eqref{eq:oddmoderelax} using
\begin{subequations}\label{eq:variational}
\begin{align}
{\cal I}_m &= \frac{1}{4} (W_0 + 3 W_1) \ln m \\
{\cal K}_m &= \frac{1}{2} (W_0 + 3 W_1) \ln m \\
{\cal S}_m &= \ \ \, (W_0-\ W_1) \ln m + 2 W_1 \bar{\Lambda}_T ,
\end{align}
\end{subequations}
such that
\begin{align}
\gamma_{m\gg1} &= \frac{\pi^3}{15} \frac{T^4}{\hbar T_F^3} \xi_m (W_0 + 3 W_1) m^4 \ln m + {\it O}(m^4) 
\end{align}
with [Eq.~\eqref{eq:xistar} with Eqs.~\eqref{eq:variational} inserted]
\begin{align}
 \xi_m = 1 - \frac{2{\cal K}_m}{2 W_1 (\bar{\Lambda}_T - 2 \ln m) + 2{\cal K}_m} .
\end{align}
The difference compared to the asymptotic form~\eqref{eq:closedformasymptotic} is the factor $\xi_m < 1$: For $W_0=W_1$, this expression evaluates to $\xi_m \simeq 1 - \ln(m^2)/\Lambda_T$, which corresponds to full dressing only for $\ln(m^2) \ll \Lambda_T$. Within the range of validity of the low-temperature expansion, however, we have $\Lambda_T - \ln(m^2) = \ln[T_F/(m^2 T)] = {\it O}(1)$ unless $T/T_F$ is exponentially small, such that the modes with $\ln (m^2) \gg 1$ are nearly rigid. Their relaxation is then well described by the rigid-mode element $L_{00}^{(m)}$.

\section{Constant interaction and numerical benchmark}\label{sec:constant}

Our results for the relaxation rate of the slowest odd mode for a constant interaction were already compared with exact diagonalization in Fig.~\ref{fig:1} in the introduction. In this section, we perform a more detailed benchmark of the odd mode as well as the complete spectrum including fast Fermi-liquid modes.

For a constant interaction potential $\tilde{V}(q) = V_0$, we have
\begin{align}
W(x,\tilde{x}) &= W_c = (g_s-1) |V_0|^2 . \label{eq:constantamplitude}
\end{align}
The angle integrals~\eqref{eq:residualcoeffs} then evaluate to $\Omega_\Delta^{(m)}[W] = \Omega_{DE}^{(m)}[W] = \Omega_\Sigma^{(m)}[W] = 0$ for odd $m$ and 
\begin{align}
    \Omega_\Delta^{(m)}[W] &= - 3 W_c \, o_m \\
    \Omega_{DE}^{(m)}[W] &= 2 W_c \, o_m \\
    \Omega_\Sigma^{(m)}[W] &= W_c \, o_m
\end{align}
for even $m$, where we evaluate
\begin{align}
    \frac{1}{4} \int_0^1 \frac{ds}{s} \, U_{m-1}^2(\sqrt{s}) = o_m 
\end{align}
in terms of the function
\begin{align}
    o_m &= \sum_{{\rm odd} \, j \leq m} \frac{1}{j} = H_m - \frac{1}{2} H_{\lfloor{m/2} \rfloor} \label{eq:defom}
\end{align}
with $H_m = \sum_{j=1}^m 1/j$ a harmonic number. At large $m$, we have $o_m = (\ln2m)/2 + \gamma_E/2 + {\it O}(1/m)$, where $\gamma_E$ is the Euler constant. Note that $o_m$ is evaluated here for even $m$, but we will later use the definition for odd $m$ as well. 
Likewise, we have
\begin{align}
\Omega_{00}^{(m)}[W] &= \frac{\pi^2 m^2}{24} 2 W_c 
\int_0^1 ds\,
\biggl[
\dfrac{T_m^2(\sqrt{s})}{s^2}
-
\dfrac{m^2}{s} 
\biggr] \nonumber \\
&= - \frac{\pi^2 m^4}{3} W_c \Bigl(o_m - p + \frac{p^2}{4}\Bigr) \\
\Omega_{01}^{(m)}[W] &= - \frac{\pi m}{4\sqrt{3}} 2 W_c
\int_0^1 ds\,
\biggl[
\dfrac{
T_m(\sqrt{s})U_{m-1}(\sqrt{s})
}{s^{3/2}}
-
\dfrac{m}{s} 
\biggr] \nonumber \\
&= \frac{\pi m^2}{\sqrt{3}} W_c \bigl(2 o_m - p\bigr) , \label{eq:Omega01const}
\end{align}
where $p=1 + 1/m$. 

\subsection{Full spectrum}

The entries of the collision matrix for $0\leq n,n' \leq 2$ read
\begin{widetext}
    \begin{align}
\frac{\hbar T_F}{T^2 W_c} L_{n'n}^{(m \, {\rm even})} &= 
\begin{pmatrix}
\dfrac{8\pi}{3} o_m & \dfrac{4\pi^2 m^2}{5\sqrt{3}} \dfrac{T}{T_F} \bigl[\bar{\Lambda}_T - 2\tilde{a}_m + 2\bigr] & \dfrac{8\pi}{3\sqrt{5}} o_m
\\
\dfrac{4\pi^2 m^2}{5\sqrt{3}} \dfrac{T}{T_F} \bigl[\bar{\Lambda}_T - 2 \tilde{a}_m + 2\bigr] & \dfrac{24\pi}{5} o_m & 0 & \ddots
\\
\dfrac{8\pi}{3\sqrt{5}} o_m & 0 & \dfrac{2\pi}{7} \bigl[12 \bar{\Lambda}_T + 5(h_3 - h)\bigr] + \dfrac{40\pi}{21} o_m
\\
& \ddots & & \ddots
\end{pmatrix} , \label{eq:Lconsteven} \\
\frac{\hbar T_F}{T^2 W_c} L_{n'n}^{(m \, {\rm odd})}
&= \begin{pmatrix}
\dfrac{2\pi^3 m^4}{15} \dfrac{T^2}{T_F^2} \Bigl[\bar{\Lambda}_T - 4\Bigl(o_m - p + \dfrac{p^2}{4}\Bigr)\Bigr]
& - \dfrac{4\pi^2 m^2}{5\sqrt{3}} \dfrac{T}{T_F} \bigl[\bar{\Lambda}_T - 4o_m + 2p\bigr] & 0
\\
- \dfrac{4\pi^2 m^2}{5\sqrt{3}} \dfrac{T}{T_F} \bigl[\bar{\Lambda}_T - 4o_m + 2p\bigr] & \dfrac{8\pi}{5} \bar{\Lambda}_T & 0 & \ddots
\\
0 & 0 & \dfrac{2\pi}{7} \bigl[12 \bar{\Lambda}_T + 5(h_3 - h)\bigr]
\\
& \ddots & & \ddots
\end{pmatrix} , \label{eq:Lconstodd}
    \end{align}
\end{widetext}
where $h_3 = \bar{M}_6/M_6 = \ln 4 - \psi(7) - \zeta'(6)/\zeta(6) \approx - 0.47$ [cf.~App.~\ref{app:logfermiintegrals}]. This form has the structure of the collision matrix discussed in Sec.~\ref{sec:asymptoticmatching}: Even modes do not have a logarithmic ${\it O}(T^2 \ln(T_F/T))$ term in the $n=0$ and $n=1$ rows or columns by particle number and energy conservation, but they do have a finite ${\it O}(T^2)$ correction. In Eq.~\eqref{eq:Lconsteven}, we also include the leading off-diagonal $01$ coefficient, which is ${\it O}(T^3)$, as for odd $m$. It requires an expansion of the integration kernel, which is done in App.~\ref{app:L01}. For odd modes, the $n=0$ row has no ${\it O}(T^2 \ln(T_F/T))$ and ${\it O}(T^2)$ terms because $\Delta_0 = E_0 = D_0 = 0$, and its scaling is set by higher-order terms. Higher odd matrix elements do not depend on $m$ since the angle integrals in  Eq.~\eqref{eq:residualcoeffs} vanish for odd $m$, a result that is specific to the constant interaction potential and does not hold in general. 

%++++++++++++++++++++++++++
\begin{figure}
	\centering
 \scalebox{1.0}{\includegraphics{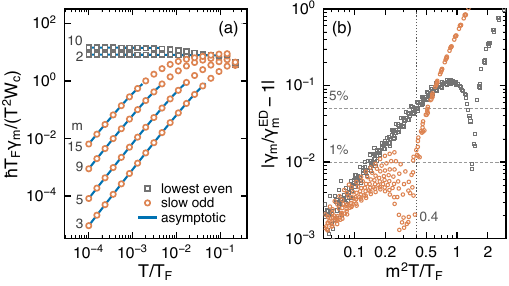}}
 \caption{(a) Lowest relaxation rates for the angular harmonics $m=2,3,5,9,10$, and $15$ as a function of temperature, obtained from exact diagonalization for a constant interaction $W_c$~\cite{hofmann23,nilsson25} with $8$ radial basis modes. Blue lines show the lowest eigenvalue of the asymptotic collision matrix of this work truncated to $8$ radial modes. (b) Relative deviation $|\gamma_m/\gamma_m^{\rm ED} - 1|$ of the lowest eigenvalue for $2 \leq m \leq 20$. Up to $m^2T/T_F = 0.4$ (dotted line), the asymptotic result  describes the lowest odd rates to within $1\%$ and the even rates to $5\%$ (dashed lines).
 }
 \label{fig:4}
\end{figure}
%++++++++++++++++++++++++++

We now benchmark the asymptotic results against the exact relaxation rates in more detail. Figure~\ref{fig:4}(a) shows the lowest odd (orange) and even (gray) relaxation rates at $m=2,10$ (even) and $m=3,5,9,15$ (odd) as a function of temperature, with the asymptotic results (obtained by diagonalizing the asymptotic collision matrix derived in this work) shown in blue. The asymptotic rates are plotted up to $m^2T/T_F = 0.4$, which we take as the boundary of applicability of the low-temperature expansion. This choice follows from Fig.~\ref{fig:4}(b), which shows the relative deviation $|\gamma_m/\gamma_m^{\rm ED} - 1|$ between the asymptotic eigenvalue $\gamma_m$ and the exact-diagonalization result $\gamma_{m}^{\rm ED}$. As is apparent from both panels, the low-temperature collision matrix accurately describes both the long-lived odd rates and the faster even-parity rates, with the boundary $m^2T/T_F=0.4$ (vertical dotted line) describing odd-parity rates up to $1\%$ accuracy and even-parity rates up to $5\%$ (horizontal lines), respectively.

\subsection{Long-lived odd modes}

For a constant potential, Eqs.~\eqref{eq:Kmcoeff},~\eqref{eq:Smcoeff}, and~\eqref{eq:Imcoeff} evaluate to
\begin{subequations} \label{eq:resultsconstant}
\begin{align}
 {\cal K}_m &= 2 W_c \bigl(2 o_m - p\bigr) , \\
 {\cal S}_m &= 2 W_c \bar{\Lambda}_T , \\
 {\cal I}_m &= 2 W_c \Bigl(o_m - \frac{p^2}{4}\Bigr) ,
\end{align}
\end{subequations}
where $o_m$ is defined in Eq.~\eqref{eq:defom} and $p$ after Eq.~\eqref{eq:Omega01const}. This gives the relative dressing
\begin{align}
\xi_m &= 1 - \frac{2(2 o_m - p)}{\bar{\Lambda}_T} \label{eq:ximconstant}
\end{align}
and the odd eigenvalue
\begin{align}
\gamma_m &= \frac{8\pi^3}{15} W_c m^4 \frac{T^4}{\hbar T_F^3}
\biggl[o_m - \frac{p^2}{4} - \frac{(2 o_m - p)^2}{\bar{\Lambda}_T}\biggr] . \label{eq:gammamconstoptimal}
\end{align}
The expression for $\xi_m$ corresponds to the dressing that minimizes the general form for $\gamma_m(\xi)$ in Eq.~\eqref{eq:oddmoderesult} for a constant potential. The rate $\gamma_m$ is its value evaluated at $\xi=\xi_m$ [Fig.~\ref{fig:1}]. The expression~\eqref{eq:gammamconstoptimal} can be evaluated in closed analytical form for all $m$. In particular, for $m=3$ we have a relative dressing
\begin{align}
\xi_{m=3} &= 1 - \frac{8}{3\bar{\Lambda}_T} \label{eq:constdressingm3}
\end{align}
and a relaxation rate
\begin{align}
    \gamma_{m=3} &= \frac{192 \pi^3}{5} W_c \frac{T^4}{\hbar T_F^3} \biggl(1 - \frac{2}{\bar{\Lambda}_T}\biggr) . \label{eq:constagammam3}
\end{align}

%++++++++++++++++++++++++++
\begin{figure}
	\centering
 \scalebox{0.9}{\includegraphics{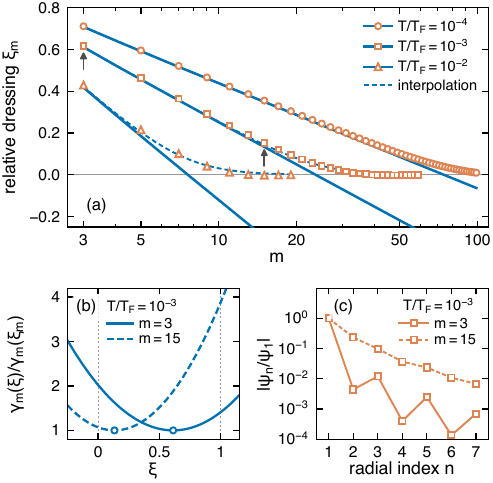}}
 \caption{(a) Relative dressing $\xi_m$ of the long-lived odd modes as a function of $m$ for  three different temperatures $T/T_F = 10^{-4}, 10^{-3}$, and $10^{-2}$. Solid blue lines: analytical low-temperature result [Eq.~\eqref{eq:ximconstant}]. Orange markers: exact-diagonalization results of Refs.~\cite{hofmann23,nilsson25}. Black arrows indicate the modes shown in panels (b) and (c). Blue dashed lines: interpolation formula~\eqref{eq:interpolation}. (b) Variational rate $\gamma_m(\xi)$ [Eq.~\eqref{eq:oddmoderesult}] as a function of $\xi$ for $m=3$ (solid) and $m=15$ (dashed) at $T/T_F=10^{-3}$. Open circles mark the minima $\xi_m$. (c) Radial components $|\psi_n/\psi_1|$ of the corresponding ED eigenvectors. Components with $n\geq 2$ are suppressed, and the mode is well described by Eq.~\eqref{eq:odddeformation} that consists of a rigid displacement of the Fermi surface and a local temperature change.}
 \label{fig:5}
\end{figure}
%++++++++++++++++++++++++++

Figure~\ref{fig:5}(a) shows the relative dressing $\xi_m$ [Eq.~\eqref{eq:ximconstant}] for three different temperatures $T/T_F=10^{-4}, 10^{-3}$, and $10^{-2}$ (blue lines) as a function of the angular index $m$ on a logarithmic scale. For comparison, we also include the result obtained from exact diagonalization at these temperatures (orange markers)~\cite{hofmann23,nilsson25}. At small $m$, they agree well, with an $m$-dependence that nearly follows the asymptotic $\ln(2m)$ scaling of $o_m$. At larger $m$, they disagree, and the full solution is suppressed toward zero, whereas the asymptotic prediction crosses zero and assumes negative values. The point at which the asymptotic description breaks down is estimated from $\xi_m=0$ with the large-$m$ asymptotics of $o_m$, which gives $m^2 T/T_F \lesssim e^{2(1-\gamma_E)+h}/4 \approx 0.55$. This is consistent with the stricter $1\%$ bound for the odd rates  [Fig.~\ref{fig:4}]. We find that the full result for $\xi_m$ beyond the asymptotic low-temperature region is accurately captured by the interpolation formula [blue dashed line in Fig.~\ref{fig:5}(a)]
\begin{align}
    \xi_m^{\rm int} &= \frac{1}{\zeta \bar{\Lambda}_T} \ln \bigl(1 + e^{\zeta \bar{\Lambda}_T \xi_m}\bigr) , \label{eq:interpolation}
\end{align}
where $\zeta = 2$ is a fit parameter and $\xi_m$ is given in Eq.~\eqref{eq:ximconstant}. This interpolation function leaves $\xi_m$ essentially unchanged for positive and moderately large values but makes it exponentially small for negative $\xi_m$.

Figure~\ref{fig:5}(b) shows the full variational form~\eqref{eq:oddmoderelax} of the relaxation rate with coefficients~\eqref{eq:resultsconstant} [equivalently, Eq.~\eqref{eq:oddmoderesult} evaluated for a constant interaction] for $m=3$ (solid line) and $m=15$ (dashed line) at $T/T_F=10^{-3}$. In both cases, the mode is partially dressed with a minimum [open circles in Fig.~\ref{fig:5}(b) and black arrows in Fig.~\ref{fig:5}(a)] at $0 < \xi_m < 1$. In particular, assuming a fully dressed mode at $m=15$ overestimates the relaxation rate by a factor of $4$. Thus, the partial thermal dressing increases the lifetime of the mode compared to the fully dressed case.

Finally, in Fig.~\ref{fig:5}(c), we show the weight of the full eigenvector from exact diagonalization at $m=3$ (solid line) and $m=15$ (dashed line) at $T/T_F=10^{-3}$ in the radial basis. The suppression of the higher radial modes $n\geq 2$ is consistent with the general argument in Sec.~\ref{sec:eigenmodelgeneral} that justifies an effective description of the low-temperature structure using the $n=0$ and $n=1$ components, i.e., in terms of a rigid deformation and a thermal dressing.

\section{Dipolar fermions}\label{sec:dipoles}

To illustrate our results for general interaction potentials, this section discusses the odd-mode relaxation rate of spin-polarized ($g_s=1$) 2D dipolar fermions. Such a setup describes a dipolar Fermi quantum gas with tight transverse confinement~\cite{lu12pra,babadi12} in which dipoles are aligned out-of-plane and interact via the repulsive side-by-side scattering component of the dipolar potential. Degenerate dipolar Fermi gases are created in experiments with ultracold quantum gases of $\mathrm{Dy}$~\cite{lu10,lu12} or $\mathrm{Er}$~\cite{aikawa14}, which have a large magnetic dipole moment, and with fermionic molecules such as $\mathrm{Na}\mathrm{K}$~\cite{schindewolf22} or $\mathrm{KRb}$~\cite{demarco19,valtolina20,lin26}. Current experiments are restricted to the quasi-2D setting at some fraction of the Fermi temperature, but the strictly 2D limit could be accessible in future studies.

The 2D dipolar interaction potential consists of a short-range part and a repulsive component that decays as $d^2/r^3$ with distance, where $d$ is the dipole moment. In Fourier space, this potential takes the form $V({\bf q}) = const. - 2 \pi d^2 |{\bf q}|$ valid for momenta $q\ell_z\ll 1$, where $\ell_z$ is the length scale that sets transverse confinement of the dipoles~\cite{fischer06,komineas07}. This corresponds to a dimensionless potential $\tilde{V}(x) = const. - 2 \lambda_d x$ with $\lambda_d = k_F \ell_d$, where $\ell_d = m^*d^2/\hbar^2$ is the dipolar length. The full potential~\eqref{eq:defW} then takes a simple form in which the constant part cancels,
\begin{align}
 W(x, \tilde{x}) &= 2 \lambda_d^2 (x-\tilde{x})^2 .
\end{align}
This form of the scattering potential fully suppresses collinear parallel processes, $W_0=0$ [for which $x \approx \tilde{x} \approx 0$, Fig.~\ref{fig:3}(c)--(d)] and enhances collinear antiparallel scattering [for which $(x,\tilde{x}) \approx (0,1)$ or $(1,0)$, Fig.~\ref{fig:3}(a) and~(f)], $W_1=2\lambda_d^2$. The effective potential that enters Eq.~\eqref{eq:oddmoderesult} takes the form
\begin{align}
W_{\rm tot}(s) &= W_{\rm ex}(1-s) + W_{\rm ho}(s) \nonumber \\
&= 2 \lambda_d^2 \Bigl[2 - s - 2 \sqrt{s(1-s)}\Bigr] .
\end{align}
The odd-mode dressing and relaxation rates are evaluated in closed form from Eqs.~\eqref{eq:Smcoeff},~\eqref{eq:Kmcoeff}, and~\eqref{eq:Imcoeff}. For the slowest nonconserved odd mode (which is $m=3$ since $m=1$ describes the conserved momentum mode), the relative dressing is
\begin{align}
    \xi_{m=3} &= 1 - \frac{1}{9} \frac{14-3\pi}{\bar{\Lambda}_T - (7+3\pi)/6} \label{eq:ximdipole}
\end{align}
and the relaxation rate
\begin{align}
\gamma_{m=3}&= \frac{4 \pi^3}{5} \lambda_d^2 \frac{T^4}{\hbar T_F^3} \Bigl[22 - 3 \pi + 3 (14-3\pi) \xi_3\Bigr] .
\end{align}
Compared with a constant interaction with the same $W_1$ [Eqs.~\eqref{eq:constdressingm3} and~\eqref{eq:constagammam3}], the relative dressing~\eqref{eq:ximdipole} remains much closer to full dressing $\xi=1$, and the relaxation rate is smaller. For example, at temperatures $T/T_F = 10^{-4}, 10^{-3}$, and $10^{-2}$, the relative dressing is $\xi_3 = 0.92, 0.88$, and $0.72$, respectively, compared to $\xi_3 = 0.71, 0.61$, and $0.41$ for the constant potential shown in Fig.~\ref{fig:5}(a), and the ratio of the constant and dipolar relaxation rates is $\gamma_{m=3}^{\rm const}/\gamma_{m=3}^{\rm dip} = 2.98, 2.76$, and $2.40$. Intuitively, both effects are expected since the dipolar interaction suppresses large-angle scattering compared to the collinear endpoint amplitude.

\section{Conclusion}\label{sec:summary}

In summary, we have presented a full leading-order calculation of the 2D Fermi-liquid collision integral for general regular isotropic potentials, including the nonlogarithmic terms beyond the temperature logarithm, and benchmarked our result against exact diagonalization for constant interactions. We find that long-lived odd-parity modes consist of a rigid Fermi-surface deformation with a concomitant local temperature modulation that reduces relaxation by collinear scattering. Since collinear scattering is only logarithmically enhanced and competes with large-angle scattering, however, the relative dressing $\xi$ is only partial and decreases with increasing angular harmonic index $m$ until the modes become nearly rigid near $m\simeq \sqrt{T_F/T}$. This result resolves a discrepancy between exact diagonalization and the leading-logarithmic prediction $m^4 \ln m$, which corresponds to full dressing instead. The dipolar interaction is an example of a potential with reduced large-angle scattering, for which the dressing is larger than for a constant interaction. 

Collision integrals are frequently evaluated to leading logarithmic order, not only for Fermi liquids~\cite{hodges71,bloom75,giuliani82,laikhtman92,menashe96,zheng96,dassarma21}, but also for graphene transport~\cite{fritz08,kashuba08}, where the expansion parameter is $1/\ln (1/\alpha(T))$ with a temperature-dependent effective fine-structure constant $\alpha(T)$. Our work provides an example where the corresponding transport lifetimes are not described accurately at this order, with a cancellation of the leading-logarithmic term (in the presence of perfect dressing $\xi=1$) and large subleading logarithmic corrections (which act to reduce the dressing). We note that a similar effect is found for transport calculations of high-temperature gauge theories, which require a resummation of terms governed by inverse logarithms of the gauge coupling $1/\ln(1/g)$~\cite{arnold00,arnold03}.

Direct solutions of the Fermi-liquid kinetic equation usually truncate the radial basis and assign a single relaxation rate per angular sector~\cite{ledwith19b,kryhin25,alvarez25,starkov26a,farrell26,benshachar26,rostami25,benshachar25a,benshachar25b,maki26,starkov26b,benshachar26b,benshachar25c,moiseenko25,thuillier26}. The closed-form collision matrix in Sec.~\ref{sec:asymptoticresults} together with the interpolation~\eqref{eq:interpolation} at higher temperatures provides both the relaxation rates and the structure of the modes for calculations beyond this approximation. In particular, the companion Letter shows that the thermal dressing should be detectable in the thermopower~\cite{hofmann26}. A natural next step is to compute transport coefficients such as the viscosity or the thermal conductivity to the same nonlogarithmic accuracy. Different from 3D Fermi liquids~\cite{brooker68,repplinger25,taillat25,taillat26,kurkjian26}, where corrections are not logarithmically suppressed and can be evaluated in closed form due to the banded structure of radial corrections, this requires the full radial matrix of Sec.~\ref{sec:asymptoticresults}, which is not banded. Here, several  distinct logarithmic effects are discussed in the literature---for example, arising from Cooper channel singularities~\cite{novikov06} or forward-scattering Coulomb singularities~\cite{menashe96,lyakhov03}---which thus far appear to lack a unified treatment.

\begin{acknowledgments}
We thank Nitay Ben-Shachar, Ulf Gran, Hadrien Kurkjian, and Carl L\"ov for discussions. This work is supported by Vetenskapsr\r{a}det (Grant No.~2024-04485), the Olle Engkvist Foundation (Grant No.~233-0339), the Knut and Alice Wallenberg Foundation (Grant No.~KAW 2024.0129), and Nordita. The symbolic evaluation of the integrals in Appendixes~\ref{app:logfermiintegrals} and~\ref{app:Fermiintegrals} was performed with the assistance of ChatGPT (OpenAI, GPT-5.6); all results were independently verified analytically using Mathematica 15.0 and by hand. We used Claude (Anthropic, Opus 5.5) to assist with Mathematica code for a numerical calculation of the collision matrix elements. The authors take full responsibility for all results.
\end{acknowledgments}

\appendix

\section{Linearized collision integral}\label{app:collisionintegral}

In this appendix, we collect basic identities for the linearized collision integral and derive Eq.~\eqref{eq:collisiondefinition}, which is the starting point of our asymptotic matching calculation. Writing $\delta f({\bf k}) = f_0({\bf k}) (1-f_0({\bf k})) h({\bf k})$ and expanding to linear order in $h$, we obtain the collision integral
\begin{widetext}
\begin{align}
&{\cal J}[\delta f({\bf k}_1)] = - \frac{2\pi}{\hbar} \int \frac{d{\bf k}_2}{(2\pi)^2} \int \frac{d{\bf k}_{1'}}{(2\pi)^2} \int \frac{d {\bf k}_{2'}}{(2\pi)^2} \, \frac{W(x, \tilde{x})}{N_0^2} f_0({\bf k}_1) f_0({\bf k}_2) \bigl(1 - f_0({\bf k}_{1'})\bigr) \bigl(1 - f_0({\bf k}_{2'})\bigr) \nonumber \\
&\quad \times \delta\bigl(\varepsilon({\bf k}_1) + \varepsilon({\bf k}_2) - \varepsilon({\bf k}_{1'}) - \varepsilon({\bf k}_{2'})\bigr) (2\pi)^2 \delta\bigl({\bf k}_1 + {\bf k}_2 - {\bf k}_{1'} - {\bf k}_{2'}\bigr)\times \Bigl[ h({\bf k}_1) + h({\bf k}_2) - h({\bf k}_{1'}) - h({\bf k}_{2'})\Bigr] . \label{eq:linearizedcollisionintegral}
\end{align}
\end{widetext}
Here, $W$ is defined in Eq.~\eqref{eq:defW} for a general scattering potential $V({\bf q})$. The integrand is symmetric under the simultaneous exchange of ${\bf k}_1$ and ${\bf k}_2$ as well as ${\bf k}_{1'}$ and ${\bf k}_{2'}$ (this symmetry corresponds to an exchange of particle labels and holds separately for each Fermi factor but only together for the interaction matrix element), and antisymmetric under a simultaneous exchange of $({\bf k}_1, {\bf k}_2)$ and $({\bf k}_{1'}, {\bf k}_{2'})$ (this exchange of initial and final states requires time-reversal symmetry of the interaction).

The matrix elements of the linearized collision integral~\eqref{eq:linearizedcollisionintegral} with respect to the inner product~\eqref{eq:innerproduct} are defined by
\begin{widetext}
\begin{align}
\langle h' | {\cal L} | h \rangle &= \frac{2\pi}{\hbar N_0 T} \int \frac{d{\bf k}_1}{(2\pi)^2} \int \frac{d{\bf k}_2}{(2\pi)^2} \int \frac{d{\bf k}_{1'}}{(2\pi)^2} \int \frac{d {\bf k}_{2'}}{(2\pi)^2} \, \frac{W(x, \tilde{x})}{N_0^2} \, 
f_0({\bf k}_1) f_0({\bf k}_2) \bigl(1 - f_0({\bf k}_{1'})\bigr) \bigl(1 - f_0({\bf k}_{2'})\bigr) \nonumber \\
&\quad\qquad \times \delta\bigl(\varepsilon({\bf k}_1) + \varepsilon({\bf k}_2) - \varepsilon({\bf k}_{1'}) - \varepsilon({\bf k}_{2'})\bigr) (2\pi)^2 \delta\bigl({\bf k}_1 + {\bf k}_2 - {\bf k}_{1'} - {\bf k}_{2'}\bigr) \nonumber \\
&\quad\qquad \times \frac{1}{4} \Bigl[ h^{\prime *}({\bf k}_1) + h^{\prime *}({\bf k}_2) - h^{\prime *}({\bf k}_{1'}) - h^{\prime *}({\bf k}_{2'})\Bigr] \Bigl[ h({\bf k}_1) + h({\bf k}_2) - h({\bf k}_{1'}) - h({\bf k}_{2'})\Bigr] . \label{eq:matrixelement}
\end{align}
\end{widetext}
To get the antisymmetry in the first square bracket in the third line, we exploit the symmetries of the integrand listed above.

We rewrite the momentum delta function in the standard way in terms of the momentum transfer using
\begin{align}
&\delta\bigl({\bf k}_1 + {\bf k}_2 - {\bf k}_{1'} - {\bf k}_{2'}\bigr) \nonumber \\
&= 
\int d{\bf q} \, 
\delta\bigl({\bf k}_1 - {\bf k}_{1'} - {\bf q}\bigr) \,
\delta\bigl({\bf k}_2 - {\bf k}_{2'} + {\bf q}\bigr) ,
\end{align}
which allows us to evaluate the integrals over ${\bf k}_{1'}$ and ${\bf k}_{2'}$ in Eq.~\eqref{eq:matrixelement}. Likewise, the energy delta function is rewritten as
\begin{align}
&\delta\bigl(\varepsilon({\bf k}_1) + \varepsilon({\bf k}_2) - \varepsilon({\bf k}_{1'}) - \varepsilon({\bf k}_{2'})\bigr) \nonumber \\ 
&= 
\hbar \int d\omega \, 
\delta\bigl(\varepsilon({\bf k}_1) - \varepsilon({\bf k}_{1'}) - \hbar \omega\bigr) \, 
\delta\bigl(\varepsilon({\bf k}_2) - \varepsilon({\bf k}_{2'}) + \hbar \omega\bigr) \nonumber \\
&= \frac{1}{\hbar v_1 v_2 q^2} \int_{-\infty}^\infty d\omega \, \delta\Bigl(\cos(\theta_1 - \theta_q) - \frac{q}{2k_1} - \frac{\omega}{v_1 q}\Bigr) \nonumber \\
&\qquad\qquad\qquad\quad \times \delta\Bigl(\cos(\theta_2 - \theta_q) + \frac{q}{2k_2} - \frac{\omega}{v_2 q}\Bigr) . \label{eq:energydelta}
\end{align}
In addition, we parametrize the remaining ${\bf k}_1$ and ${\bf k}_2$ integrals using polar coordinates,
\begin{align}
\int d{\bf k} &= \frac{m^*}{\hbar^2 \beta} \int_{-\beta\mu}^\infty dw_k \int_{-\pi}^{\pi} d\theta_k .
\end{align}
The energy-split delta functions~\eqref{eq:energydelta} restrict the angles $\theta_1$ and $\theta_2$ to the scattering angles in Eq.~\eqref{eq:costheta}. With this parametrization, the Fermi factor reads
\begin{align}
F_{121'2'} &= 
\frac{1}{16 \cosh \frac{w_1}{2} \, \cosh \frac{w_2}{2} \, \cosh \frac{w_{1'}}{2} \, \cosh \frac{w_{2'}}{2}} \label{eq:fermifactors} .
\end{align}
By energy conservation, the Fermi factor is symmetric under permutation of the four states, and it depends only on three independent energy parameters, for example, $w_1$ and $w_2$ as well as the transferred energy $w$.

The matrix element~\eqref{eq:matrixelement} becomes
\begin{widetext}
\begin{align}
L_{n'n}^{(m)} &= 
\frac{T^2}{\hbar T_F} 
\biggl\langle \frac{1}{32\pi^2} 
\int_{-\pi}^\pi d\theta_q \, \sum_{\theta_1^\star, \theta_2^\star} \int_0^\infty \frac{dx}{x} \, \frac{v_F^2}{v_1 v_2} \frac{W(x, \tilde{x})}{|\sin \theta_1^\star \sin \theta_2^\star \, |} \, \biggl({\sum_{i}}^\prime u_{n'}({\bf k}_i) e^{-im\theta_{i}}\biggr) \biggl({\sum_{j}}^\prime u_n({\bf k}_j) e^{im\theta_j}\biggr) \biggr\rangle_F , \label{eq:Lnpnm1st}
\end{align}
\end{widetext}
where $\langle \ldots \rangle_F$ is defined in Eq.~\eqref{eq:Fermiintegral}, and $\theta_1^\star$ and $\theta_2^\star$ run over each zero of Eqs.~\eqref{eq:costhetat1} and~\eqref{eq:costhetat2} (cf.~also Fig.~\ref{fig:2}), and we define the shorthand
\begin{align}
{\sum_{j}}^\prime u_n({\bf k}_j) e^{im\theta_j} &= 
u_n({\bf k}_1) e^{im\theta_1} + u_n({\bf k}_2) e^{im\theta_2} \nonumber \\
&\ - u_n({\bf k}_{1'}) e^{im\theta_{1'}} - u_n({\bf k}_{2'}) e^{im\theta_{2'}} .
\end{align}
The integral over $\theta_q$ in Eq.~\eqref{eq:Lnpnm1st} is performed trivially by rotational invariance: If the basis functions share the same angular harmonic index $m$, it contributes a factor of $2\pi$, otherwise the integral is zero. Note that with the integration variables $w_1, w_2, w$, and $x$, the parameter $\tilde{x}$ in $W$ is not an independent variable but determined by the constraint Eq.~\eqref{eq:constraint}.

\section{Radial basis functions}\label{app:basisfunctions}

In this appendix, we collect identities for the low-temperature radial basis functions $u_n(w)$ that are used to diagonalize the linearized collision integral~\cite{nilsson25}. The radial functions form an orthonormal set with respect to the inner product
\begin{align}
\langle u_{n'}^{} | u_{n}^{} \rangle &= \int_{-\infty}^\infty dw \, \frac{u_{n'}(w) u_n(w)}{4 \cosh^2(w/2)} .
\end{align}
The initial basis functions are $u_0(w) \sim 1$ and $u_1(w) \sim w$, from which higher-order polynomials are generated using a Gram-Schmidt procedure. The first basis functions are~\cite{hofmann23,nilsson25}
\begin{align}
u_0(w) &= 1 \\
u_1(w) &= \frac{\sqrt{3}}{\pi } w \label{eq:u1} \\
u_2(w) &= \frac{3 \sqrt{5}}{4 \pi ^2} w^2-\frac{\sqrt{5}}{4} \\ 
u_3(w) &= \frac{5 \sqrt{7}}{12 \pi ^3} w^3-\frac{7 \sqrt{7}}{12 \pi } w 
\end{align}
and so on. We note a recurrence relation between low-temperature basis functions at different orders:
\begin{align}
w \, u_{n}(w) &= \rho_{n+1} \, u_{n+1}(w) + \rho_n \, u_{n-1}(w) \label{eq:jacobi}
\end{align}
with
\begin{align}
\rho_n &= \frac{\pi n^2}{\sqrt{(2n-1) (2n+1)}} .
\end{align}
From it, we infer the leading monomial coefficients of the orthogonal polynomials $u_n$. To this end, write 
\begin{align}
u_n(w) = \kappa_n (w^n + \nu_n w^{n-2} + \ldots) \label{eq:expansion}
\end{align}
and match terms on both sides of Eq.~\eqref{eq:jacobi}, which gives $\kappa_{n+1} = \kappa_{n}/\rho_{n+1}$ and $\kappa_n \nu_n = \kappa_{n-1} \rho_n + \kappa_{n+1} \rho_{n+1} \nu_{n+1}$. The latter identity is rewritten as $\nu_{n+1} = \nu_n - \kappa_{n-1} \rho_n/\kappa_n = \nu_n - \rho_n^2$, with initial condition $\nu_1 = 0$, which is iterated to give 
\begin{align}
\nu_n = - \sum_{j=1}^{n-1} \rho_j^2 = - \frac{\pi^2}{6} \frac{n (n-1) [n(n-1)+1]}{2n-1} .
\end{align}

%++++++++++++++++++++++++++
\begin{figure}
	\centering
 \includegraphics{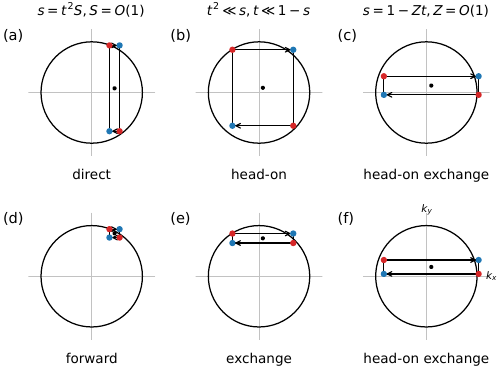}
 \caption{Sketch of different scattering configurations that contribute to the matched asymptotic expansion of the collision integral at low temperatures, where we abbreviate $t=T/T_F$. Different columns describe the three different integration regions  which match at an intermediate overlap scale. Top and bottom rows describe processes at fixed momentum transfer with scattering configurations in the opposite and equal momentum half-planes.}
 \label{fig:6}
\end{figure}
%++++++++++++++++++++++++++

\section{Asymptotic scattering configurations}\label{app:scatteringconfigurations}

In this appendix, we discuss the different scattering configurations that enter the three integrals~\eqref{eq:matching} in the asymptotic integration regions. We illustrate individual processes in Fig.~\ref{fig:6}, where columns describe the inner, intermediate, and outer integration region, and rows describe opposite (top) and equal (bottom) half-plane scattering. 

In the inner region, the momentum transfer is of the order of the lower integration boundary $l_{\rm max}$ [Eq.~\eqref{eq:lowerboundary}], i.e., $s=x^2={\it O}((T/T_F)^2)$. Equivalently, processes are described by the scaling variable $s = S (T/T_F)^2$ with $S = {\it O}(1)$. Opposite half-plane scattering describes direct processes [Fig.~\ref{fig:6}(a)], and equal half-plane scattering describes forward processes [Fig.~\ref{fig:6}(d)]. In the intermediate region, we have $s = {\it O}(T^0)$ with the restriction $(T/T_F)^2 \ll s$ and $1-s \gg T/T_F$. Here, opposite half-plane scattering describes head-on processes [Fig.~\ref{fig:6}(b)] and equal half-plane scattering describes exchange processes [Fig.~\ref{fig:6}(e)]. The outer  region is set by the difference of the two upper roots~\eqref{eq:upperboundary} and extends over a larger $s$-range ${\it O}(T/T_F)$ than the inner region. In this outer region, we parametrize $s = 1 - (T/T_F) Z$ with $Z = {\it O}(1)$. Scattering in both equal and opposite half-planes describes head-on exchange processes [Figs.~\ref{fig:6}(c) and~(f)].

Quantitatively, the low-temperature expansion of the angular factors~\eqref{eq:costheta} in these regions reads
\begin{widetext}
\begin{center}
\begin{tabular}{llll}
& $s = S (T/T_F)^2, \ S = {\it O}(1)$ & $(T/T_F)^2 \ll s, \ 1-s \gg T/T_F$ & $s = 1 - Z (T/T_F), \ Z = {\it O}(1)$\\
\hline \\[-1ex]
	$\cos \theta_1^\star$ \qquad\quad & $\dfrac{w}{4\sqrt{S}} + \Bigl(+ \sqrt{S} - \dfrac{w w_1}{8\sqrt{S}}\Bigr) \dfrac{T}{T_F}$ \qquad \quad
&
	$+ \sqrt{s} + \Bigl(\dfrac{w}{4\sqrt{s}} - \dfrac{w_1 \sqrt{s}}{2}\Bigr) \dfrac{T}{T_F}$ \qquad\quad 
&
	$+ 1 + \Bigl(\dfrac{w}{4} - \dfrac{Z}{2} - \dfrac{w_1}{2}\Bigr) \dfrac{T}{T_F}$ 
\\[3ex]
	$\cos \theta_2^\star$ & $\dfrac{w}{4\sqrt{S}} + \Bigl(- \sqrt{S} - \dfrac{w w_2}{8\sqrt{S}} \Bigr) \dfrac{T}{T_F}$ \qquad\quad
&
	$- \sqrt{s} + \Bigl(\dfrac{w}{4\sqrt{s}} + \dfrac{w_2 \sqrt{s}}{2}\Bigr) \dfrac{T}{T_F}$ \qquad\quad 
&
	$- 1 + \Bigl(\dfrac{w}{4} + \dfrac{Z}{2} + \dfrac{w_2}{2}\Bigr) \dfrac{T}{T_F}$ 
\\[3ex]
	$\cos \theta_{1'}^\star$ & $\dfrac{w}{4\sqrt{S}} + \biggl(- \sqrt{S} - \dfrac{w w_{1'}}{8\sqrt{S}} \biggr) \dfrac{T}{T_F}$ \qquad\quad
&
	$- \sqrt{s} + \Bigl(\dfrac{w}{4\sqrt{s}} + \dfrac{w_{1'} \sqrt{s}}{2}\Bigr) \dfrac{T}{T_F}$ \qquad\quad 
&
	$- 1 + \Bigl(\dfrac{w}{4} + \dfrac{Z}{2} + \dfrac{w_{1'}}{2}\Bigr) \dfrac{T}{T_F}$ 
\\[3ex]
	$\cos \theta_{2'}^\star$ & $\dfrac{w}{4\sqrt{S}} + \Bigl(+ \sqrt{S} - \dfrac{w w_{2'}}{8\sqrt{S}}\Bigr) \dfrac{T}{T_F}$ \qquad\quad
&
	$+ \sqrt{s} + \Bigl(\dfrac{w}{4\sqrt{s}} - \dfrac{w_{2'} \sqrt{s}}{2}\Bigr) \dfrac{T}{T_F}$ \qquad\quad 
&
	$+ 1 + \Bigl(\dfrac{w}{4} - \dfrac{Z}{2} - \dfrac{w_{2'}}{2}\Bigr) \dfrac{T}{T_F}$ 
\end{tabular}

\end{center}
\end{widetext}
Noting that the last terms in parentheses follow by expanding the velocity ratio $v_\alpha/v_F = \sqrt{1 + w_\alpha T/T_F} = 1 + w_\alpha T/(2T_F) + \ldots$ at this order, these are simple expressions: All processes describe scattering with a center-of-mass component $K_x/k_F = wT/(4xT_F)$ [cf.~Eq.~\eqref{eq:commomentum}] and a momentum transfer $q=2k_F\sqrt{s}$. The scattering configurations merge in the overlap regions $(T/T_F)^2 \ll \delta \ll 1$ and $T/T_F \ll \delta' \ll 1$.

\section{Asymptotic kernels}\label{app:asymptotickernels}

This appendix collects the asymptotic kernels that enter Eq.~\eqref{eq:matching}. These kernels are obtained by expanding $\cos m \theta$ and $\sin m \theta$ in Eq.~\eqref{eq:CnSn} to leading order in $T/T_F$. To perform this expansion, we follow Ref.~\cite{ledwith19} and expand Eq.~\eqref{eq:costheta} to leading order in temperature making use of the following Chebyshev identities: First, for a given $\cos \theta^\star = v + \delta v$ with $|v|<1$ (which applies for the inner and intermediate regions), we have 
\begin{align}
\cos m \theta^\star &= T_m(v) + m U_{m-1}(v) \delta v + {\it O}(\delta v^2) \label{eq:Tmexpansion} \\
\sin m \theta^\star &= \sqrt{1-v^2} \, U_{m-1}(v) - \frac{m T_m(v)}{\sqrt{1-v^2}} \delta v + {\it O}(\delta v^2) ,
\end{align}
with $T_m(-v) = (-1)^m T_m(v)$ and $U_{m-1}(-v) = (-1)^{m-1} U_{m-1}(v)$, i.e., $T_m$ changes sign for odd $m$ while $U_{m-1}$ does not, and vice versa. Due to the factor $m$ in the next-to-leading order term of Eq.~\eqref{eq:Tmexpansion} and $\delta v = {\it O}(T/T_F)$, this expansion requires $mT/T_F \ll 1$, which implies the constraint $m\ll T_F/T$ on the angular sectors that are described in this expansion. Second, in the outer region where $\cos \theta = \pm (1 - \delta v)$, we use
\begin{align}
\cos m \theta &= (\pm 1)^m (1 - m^2 \delta v) + {\it O}(\delta v^2) \label{eq:Tmexpansion2} \\
\sin m \theta &= (\pm 1)^{m+1} m \sqrt{2 \delta v} + {\it O}(\delta v^{3/2}) .
\end{align}
This endpoint expansion now implies the stronger constraint $m^2 T/T_F \ll 1$ or $m\ll \sqrt{T_F/T}$. This constraint sets the regime of applicability of the low-temperature expansion, and it implies the weaker constraint on the expansion of the inner region above. We obtain for the kernel in Eq.~\eqref{eq:matching}:

(i) for $s = S (T/T_F)^2, \ S = {\it O}(1)$,
\begin{align}
K_{n'n}^{(m)+}(s) &= 2 \Delta_{n'} \Delta_n \label{eq:kernelforward}
\end{align}
and
\begin{align}
K_{n'n}^{(m)-}(s) &= 2 D_{n'} D_n + 2 \, \bigl( \Delta_{n'} \Delta_n - D_{n'} D_n\bigr) \, T_m^2\Bigl(\dfrac{w}{4\sqrt{S}}\Bigr) . \label{eq:Kminner}
\end{align}
The argument of the Chebyshev polynomial is $w/(4\sqrt{S}) = K/k_F$ in this region, or $\sqrt{1-\tilde{x}^2} = \sqrt{1- \tilde{s}}$.

(ii) for $(T/T_F)^2 \ll s, \ 1-s \gg T/T_F$,
\begin{align}
&K_{n'n}^{(m)+}(s) \nonumber \\
&= 
\begin{cases}
2 E_{n'} E_n + 2 \, \bigl( \Delta_{n'} \Delta_n - E_{n'} E_n\bigr) \, T_m^2\bigl(\sqrt{s}\bigr) & m \ {\rm even} \\
2 \Delta_{n'} \Delta_n + 2 \, \bigl( E_{n'} E_n - \Delta_{n'} \Delta_n \bigr) \, T_m^2\bigl(\sqrt{s}\bigr) & m \ {\rm odd} ,
\end{cases} 
\end{align}
and
\begin{align}
&K_{n'n}^{(m)-}(s) \nonumber \\
&= 
\begin{cases}
2 \Sigma_{n'} \Sigma_n + 2 \, \bigl( \Delta_{n'} \Delta_n - \Sigma_{n'} \Sigma_n\bigr) \, T_m^2\bigl(\sqrt{s}\bigr) & m \ {\rm even} \\
2 D_{n'} D_n + 2 \, \bigl( E_{n'} E_n - D_{n'} D_n\bigr) \, T_m^2\bigl(\sqrt{s}\bigr) & m \ {\rm odd} 
\end{cases} \label{eq:Knpninerminus}
\end{align}

(iii) for $s = 1 - Z (T/T_F), \ Z = {\it O}(1)$,
\begin{align}
K_{n'n}^{(m)\sigma} (s) &= 
\begin{cases}
2 \, \Delta_{n'} \Delta_n & m \ {\rm even} \\
2 \, E_{n'} E_n & m \ {\rm odd} 
\end{cases} \label{eq:kernelheadonexchange}
\end{align}
for both branches $\sigma = \pm$.

Since $\Delta_0=E_0=D_0=0$ but $\Sigma_0=4$, for odd $m$ these leading-order kernels vanish identically if $n'=0$ or $n=0$, and we need to proceed to the next order with higher corrections in temperature. Keeping the next nonvanishing order in the Chebyshev expansion for odd $m$ and $n'=0$ gives:

(i) for $s = S (T/T_F)^2, \ S = {\it O}(1)$
\begin{align}
K_{00}^{(m)+}(s) &= 0 \\
K_{0n}^{(m)+}(s) &= 0 ,
\end{align}
and
\begin{align}
K_{00}^{(m)-}(s) &= 32 m^2 \biggl(\dfrac{T}{T_F}\biggr)^2 S \biggl[1-\biggl(\frac{w}{4\sqrt{S}} \biggr)^2\biggr] T_m^2\biggl(\frac{w}{4\sqrt{S}} \biggr) , \\
K_{0n}^{(m)-}(s) &= - 8 m \dfrac{T}{T_F} \sqrt{S} \, D_n \biggl[1-\biggl(\frac{w}{4\sqrt{S}} \biggr)^2\biggr] \nonumber \\
&\qquad \times T_m\biggl(\frac{w}{4\sqrt{S}} \biggr) U_{m-1}\biggl(\frac{w}{4\sqrt{S}} \biggr) ,
\end{align}

(ii) for $(T/T_F)^2 \ll s, \ 1-s \gg T/T_F$,
\begin{align}
K_{00}^{(m)+}(s) &= 2 m^4 \tilde{w}^2 \frac{T^2}{T_F^2} s \frac{U_{m-1}^2\bigl(\sqrt{s}\bigr)}{m^2} , \\
K_{0n}^{(m)+}(s) &= - 2 m^2 \frac{T}{T_F} \tilde{w} \sqrt{s} E_n \, T_m\bigl(\sqrt{s}\bigr) \frac{U_{m-1}\bigl(\sqrt{s}\bigr)}{m} 
\end{align}
and
\begin{align}
K_{00}^{(m)-}(s) &= 2 m^4 \tilde{w}^2 \frac{T^2}{T_F^2} s \, \frac{U_{m-1}^2\bigl(\sqrt{s}\bigr)}{m^2} \nonumber \\
&\qquad + 2 m^2 w^2 \frac{T^2}{T_F^2} \frac{1-s}{s} \, T_m^2\bigl(\sqrt{s}\bigr) , \\
K_{0n}^{(m)-}(s) &= - 2 m^2 \frac{T}{T_F} \biggl[\tilde{w} \sqrt{s} E_n + w \frac{1-s}{\sqrt{s}} D_n \biggr] \, \nonumber \\
&\qquad \times T_m\bigl(\sqrt{s}\bigr) \frac{U_{m-1}\bigl(\sqrt{s}\bigr)}{m} . 
\end{align}

(iii) for $s = 1 - Z (T/T_F), \ Z = {\it O}(1)$
\begin{align}
K_{00}^{(m)\pm}(s) &= 2 m^4 \frac{T^2}{T_F^2} \tilde{w}^2 , \\
K_{0n}^{(m)\pm}(s) &= - 2 m^2 \frac{T}{T_F} \tilde{w} E_n .
\end{align}

\section{Even $m$ off-diagonal collision matrix element $L_{01}^{(m)}$}\label{app:L01}

In this appendix, we evaluate the collision matrix element $L_{01}^{(m)}$ for even $m$. This requires an expansion beyond the order described in the main text, and is needed to describe the thermal dressing of even modes.

\begin{figure}
	\centering
 \scalebox{0.9}{\includegraphics{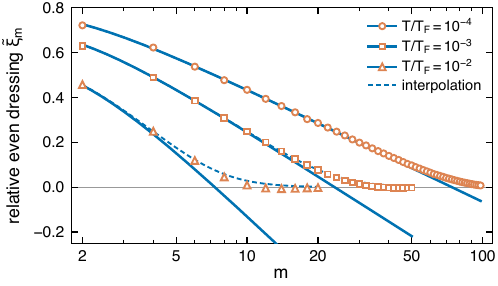}}
 \caption{Relative even-mode dressing $\tilde{\xi}_m$ at three different temperatures $T/T_F = 10^{-4}, 10^{-3}$, and $10^{-2}$ as a function of the angular harmonic index $m$. Blue lines are results of the analytical low-temperature expansion~\eqref{eq:tildexim}, orange markers are extracted from the exact-diagonalization results of Refs.~\cite{hofmann23,nilsson25}. Dashed line: Interpolation~\eqref{eq:interpolation} with $\xi_m\to \tilde{\xi}_m$.}
 \label{fig:7}
\end{figure}

By Eqs.~\eqref{eq:collisionmatrix}--\eqref{eq:defB}, the $T^2$ term and its logarithmic correction vanish after performing the Fermi average. The only nonzero contribution at this order before averaging stems from Eq.~\eqref{eq:Knpninerminus} since only $\Sigma_0 \neq 0$:
\begin{align}
K_{01}^{(m)-}(s) &= 
2 \Sigma_{0} \Sigma_1 \bigl( 1 -  T_m^2(\sqrt{s})\bigr) \nonumber \\
&= \frac{16\sqrt{3}}{\pi} (w_1+w_2) \bigl( 1 -  T_m^2(\sqrt{s})\bigr) , \label{eq:oddL01contrib}
\end{align}
which is odd in the energy and vanishes when taking the average~\eqref{eq:Fermiintegral}. Pushing the expansion to the next order in temperature gives two contributions: First, the kernel needs to be expanded in the same way as described for the odd-$m$ elements $L_{00}$ and $L_{01}$ in App.~\ref{app:asymptotickernels}. Second, there is an additional linear temperature correction that follows by expanding the upper roots Eq.~\eqref{eq:upperboundary} [the lower roots are of higher order in temperature and do not affect the result at this order]. Collecting terms in this expansion in the same way as described for other coefficients gives the result
\begin{align}
L_{01}^{(m)} &= c_m \Phi_{11}^{DE}\, \frac{T^3}{\hbar T_F^2} \Bigl( W_1 \bar{\Lambda}_T - 2 \tilde{A}_m \Bigr)
\end{align}
with
\begin{align}
\tilde{A}_m[W] &= W_1 (\tilde{a}_m - 1) \nonumber\\
&+ \frac{1}{4} \int_0^1 ds\, U_{m-1}^2(\sqrt{s}) \biggl\{ \bar{\xi}(s)\, \frac{W_{\rm tot}(s) - 2W_1}{s} \nonumber\\
&
+ \Bigl(\frac{1}{m^2 s} - \bar{\xi}(s)\Bigr) \frac{W_{\rm ho}(s) - W_1}{1-s} - \frac{1}{m^2 s} \partial_1 W_{\rm ho}(s) \biggr\} , 
\label{eq:Atilde}
\end{align}
where $\partial_1 W(x, \tilde{x}) = \partial W/\partial (x^2)|_{\tilde{x}}$ and 
\begin{align}
\tilde{a}_m &= \Bigl(2 + \frac{1}{m^2}\Bigr) o_m .
\label{eq:atilde}
\end{align}
The first line of Eq.~\eqref{eq:Atilde} is the full result for a constant interaction. 
We define the relative even-mode dressing $\tilde{\xi}_m$ by writing 
\begin{align}
L_{01}^{(m)} = c_m \Phi_{11}^{DE}\, \frac{T^3}{\hbar T_F^2} W_1 \bar{\Lambda}_T\, \tilde{\xi}_m . \label{eq:tildexim}
\end{align}
For a constant interaction, $\tilde{\xi}_m = 1 - 2 (\tilde{a}_m-1)/\bar{\Lambda}_T$. 
Figure~\ref{fig:7} compares results for a constant interaction with the exact-diagonalization data of Refs.~\cite{hofmann23,nilsson25}, from which $\tilde{\xi}_m$ is extracted as in Eq.~\eqref{eq:tildexim}. We find agreement with the analytical low-temperature form in the same parameter range as for odd modes [Fig.~\ref{fig:5}], and the same interpolation formula~\eqref{eq:interpolation} (with $\xi_m$ replaced by $\tilde{\xi}_m$) describes the data beyond the low-temperature region.

%++++++++++++++++++++++++++
%++++++++++++++++++++++++++

\section{Generating function for Fermi factor integrals}\label{app:logfermiintegrals}

In this appendix, we evaluate the Fermi energy integrals in Eq.~\eqref{eq:resFHln}. Inspired by the discussion by~\citet{ledwith19}, we consider a generating function that we extend to include the logarithmic frequency integrals:
\begin{align}
&G_\sigma(\lambda,\rho) = \int_{-\infty}^\infty dw \, 
\begin{Bmatrix}
1 \\
\ln \dfrac{4}{|w|}
\end{Bmatrix}
\nonumber \\
&\qquad \times
\int_{-\infty}^\infty dw_1 \, f_0(w_1) [1-f_0(w_1-w)] \nonumber \\
&\qquad \times
\int_{-\infty}^\infty dw_2 \, f_0(w_2) [1-f_0(w_2+w)] 
\nonumber \\
&\qquad \times
\bigl(P(\lambda) + \sigma Q(\lambda)\bigr) 
\bigl(P(\rho) + \sigma Q(\rho)\bigr)
 ,
\end{align}
where $P(\lambda) = e^{\lambda w_1} - e^{\lambda (w_1 - w)}$ and $Q(\lambda) = e^{\lambda w_2} - e^{\lambda (w_2 + w)}$. Using
\begin{align}
\int_{-\infty}^\infty dw_1 f_0(w_1) [1-f_0(w_1-w)] e^{\lambda w_1} &= \frac{\pi}{\sin \pi \lambda} \frac{e^{\lambda w} - 1}{e^w - 1} ,
\end{align}
this is evaluated as 
\begin{align}
G_\sigma(\lambda,\rho) &= \int_{-\infty}^\infty dw \, \begin{Bmatrix}
1 \\
\ln \dfrac{4}{|w|}
\end{Bmatrix} \times \frac{w^2}{4 \sinh^2 w/2} R_\sigma(\lambda, \rho, w)
\end{align}
with
\begin{align}
R_\sigma &= \frac{16}{w^2} \biggl[\frac{\pi w}{\sin \pi (\lambda + \rho)} \sinh \frac{\lambda w}{2} \sinh \frac{\rho w}{2} \sinh \frac{(\lambda + \rho) w}{2} \nonumber \\
&\quad - \frac{2\sigma \pi^2}{\sin \pi \lambda \sin \pi \rho} \sinh^2 \frac{\lambda w}{2} \sinh^2 \frac{\rho w}{2}
\biggr] .
\end{align}
We now define the coefficients
\begin{align}
R_{n'n}^\sigma(w) &= u_{n'}(\partial_\lambda) u_n(\partial_\rho) R_\sigma(\lambda, \rho, w) \Bigr|_{\lambda = \rho = 0} ,
\end{align}
such that
\begin{alignat}{4}
\bar{\Phi}_{n'n}^\Delta &= \frac{1}{4\pi} \int_{-\infty}^\infty dw \, \frac{w^2}{4 \sinh^2 w/2} R_{n'n}^+(w) \ln \frac{4}{|w|} \\
\Phi_{n'n}^\Delta &= \frac{1}{4\pi} \int_{-\infty}^\infty dw \, \frac{w^2}{4 \sinh^2 w/2} R_{n'n}^+(w) \\
\bar{\Phi}_{n'n}^{DE} &= \frac{1}{4\pi} \int_{-\infty}^\infty dw \, \frac{w^2}{4 \sinh^2 w/2} R_{n'n}^-(w) \ln \frac{4}{|w|}\\
\Phi_{n'n}^{DE} &= \frac{1}{4\pi} \int_{-\infty}^\infty dw \, \frac{w^2}{4 \sinh^2 w/2} R_{n'n}^-(w) .
\end{alignat}
The parity symmetry of the integrand implies that $\bar{\Phi}_{n'n}^\Delta$ and $\bar{\Phi}_{n'n}^{DE}$ as well as $\Phi_{n'n}^\Delta$ and $\Phi_{n'n}^{DE}$ are only nonzero for $n+n'$ even. Formally, we expand $R_{n'n}^\sigma$ as an even polynomial in $w$
\begin{align}
R_{n'n}^\sigma &= \sum_{k=1}^{(n+n')/2} p_{n'n,k}^\sigma w^{2k} ,
\end{align}
such that
\begin{alignat}{4}
\bar{\Phi}_{n'n}^\Delta &= \frac{1}{4\pi} \sum_{k=1}^{(n+n')/2} p_{n'n,k}^+ \bar{M}_{2k+2} \\
\Phi_{n'n}^\Delta &= \frac{1}{4\pi} \sum_{k=1}^{(n+n')/2} p_{n'n,k}^+ M_{2k+2} , \\
\bar{\Phi}_{n'n}^{DE} &= \frac{1}{4\pi} \sum_{k=1}^{(n+n')/2} p_{n'n,k}^- \bar{M}_{2k+2} \\
\Phi_{n'n}^{DE} &= \frac{1}{4\pi} \sum_{k=1}^{(n+n')/2} p_{n'n,k}^- M_{2k+2} ,
\end{alignat}
where
\begin{align}
M_{2r} &= \int_{-\infty}^\infty dw \, \frac{w^{2r}}{4 \sinh^2 w/2} = 2 \sum_{k=1}^\infty k \int_{0}^\infty dw \, w^{2r} e^{-kw} \nonumber \\
&= 2 \sum_{k=1}^\infty k \frac{\Gamma(2r+1)}{k^{2r+1}} = 2 \Gamma(2r+1) \zeta(2r) 
\end{align}
and
\begin{align}
\bar{M}_{2r} &= \int_{-\infty}^\infty dw \, \frac{w^{2r}}{4 \sinh^2 w/2} \ln \frac{4}{|w|} \nonumber \\
&= M_{2r} h_{r} \label{eq:hatM}
\end{align}
with
\begin{align}
    h_{r} &= \ln 4 - \psi(2r+1) - \frac{\zeta'(2r)}{\zeta(2r)} .  \label{eq:defhr}
\end{align}
Note that $h\equiv h_2$ in the main text. 
Equation~\eqref{eq:hatM} follows using
\begin{align}
\bar{M}_{2r} &= \lim_{\alpha\to 0} \frac{\partial}{\partial \alpha} \int_{-\infty}^\infty dw \, \frac{w^{2r}}{4 \sinh^2 w/2} \biggl(\frac{4}{|w|}\biggr)^\alpha \nonumber \\
&= M_{2r} \, \ln 4 + \lim_{\alpha\to 0} \frac{\partial}{\partial \alpha} M_{2r-\alpha} .
\end{align}

\section{Banded Fermi integrals}\label{app:Fermiintegrals}

The coefficient matrix of the logarithmic terms $\Phi_{n'n}^\Delta$ and $\Phi_{n'n}^{DE}$ can also be evaluated in closed form without using the generating function in App.~\ref{app:logfermiintegrals}. The result is
\begin{subequations}\label{eq:resFH}
\begin{alignat}{4}
\Phi^{\Delta}_{n,n-2} &= \frac{\alpha_{n-2}^+}{\pi} &\qquad& \Phi^{\Delta}_{nn} &= \frac{\beta_n^+}{\pi} &\qquad& \Phi^{\Delta}_{n,n+2} &= \frac{\alpha_n^+}{\pi} , \\
\Phi^{DE}_{n,n-2} &= \frac{\alpha_{n-2}^-}{\pi} &\qquad& \Phi^{DE}_{nn} &= \frac{\beta_n^-}{\pi} &\qquad& \Phi^{DE}_{n,n+2} &= \frac{\alpha_n^-}{\pi} , \label{eq:appH}
\end{alignat}
\end{subequations}
with
\begin{align}
\alpha_n^\pm 
&= \frac{\pi^2 (n+1) (n+2) [n (n+3) \pm 2 [(-1)^n-1]]}{2 (2n+3) \sqrt{(2n+1) (2n+5)}} \label{eq:resA} \\
\beta_n^\pm 
&= \frac{\pi^2 n^2 (n+1)^2}{(2n-1) (2n+3)} \pm [(-1)^n - 1] \frac{2 \pi^2 [n (n+1)-1]}{(2n-1) (2n+3)} . \label{eq:resC}
\end{align}
The starting point for a derivation of this result is the form
\begin{align}
\Phi_{n'n}^\Delta &= 
\frac{1}{\pi} \int_{-\infty}^\infty dw_1 \, \frac{u_{n'}(w_1)}{4 \cosh^2 w_1/2} {\cal T}_+[u_n]
= \frac{1}{\pi} \langle u_{n'} | {\cal T}_+[u_n] \rangle \label{eq:defF} \\
\Phi_{n'n}^{DE} &= 
\frac{1}{\pi} \int_{-\infty}^\infty dw_1 \, \frac{u_{n'}(w_1)}{4 \cosh^2 w_1/2} {\cal T}_-[u_n]
= \frac{1}{\pi} \langle u_{n'} | {\cal T}_-[u_n] \rangle \label{eq:defH}
\end{align}
with
\begin{align}
{\cal T}_\pm[u] &= 
\int_{-\infty}^\infty dw_2 \int_{-\infty}^\infty dw \, \frac{\cosh^2 \frac{w_1}{2}}{\bigl(\cosh \frac{w_1+w_2}{2} + \cosh \frac{w_1-w_2}{2}\bigr)} \nonumber \\
&\quad \times \frac{u(w_1) \pm u(w_2) - u(w_{1'}) \mp u(w_{2'})}{\bigl(\cosh \frac{w_1+w_2}{2} + \cosh (\frac{w_1-w_2}{2} - w)\bigr)}. 
\end{align}
We change integration variables to $U = e^{w_{1'}} = e^{w_1-w}$ and $V = e^{w_{2'}} = e^{w_2+w}$ with Jacobian $dw_2 dw = dU dV/(UV)$. Defining $X=e^{w_1}$ gives
\begin{align}
{\cal T}_\pm[u] &= 
\int_{0}^\infty dU \int_{0}^\infty dV \, \frac{(1+X)}{(1+U) (1+V) (X + UV)} \nonumber \\
&\times \Bigl(u(\ln X) \pm u(\ln \tfrac{UV}{X}) - u(\ln U) \mp u(\ln V)\Bigr) .
\end{align}
Consider the generating function $u(w) = e^{\lambda w}$. This gives
\begin{align}
{\cal T}_\pm[e^{\lambda w}] &= 
\int_{0}^\infty dU \int_{0}^\infty dV \, \frac{(1+X)}{(1+U) (1+V) (X + UV)} \nonumber \\
&\quad\times\biggl(X^\lambda \pm \frac{U^\lambda V^\lambda}{X^\lambda} - U^\lambda \mp V^\lambda\biggr) \nonumber \\
&=\frac{\pi^2 + w^2}{2} e^{\lambda w} + (1\pm2) \frac{\pi w}{\sin \pi \lambda} \nonumber \\
&\quad+ \frac{\pi^2}{\sin^2 \pi \lambda} \Bigl[\pm e^{- \lambda w} - (1 \pm 1) e^{\lambda w} + \cos \pi \lambda\Bigr] .
\end{align}
From this form, we infer that ${\cal T}_{\pm}[w^n]$ generates a polynomial of degree at most $n+2$ (terms of order $\lambda^n$ on the right-hand side contain at most the monomial $w^{n+2}$), and that it preserves the parity of a polynomial. Together with the Hermitian property $\langle {\cal T}_\pm p | q \rangle = \langle p | {\cal T}_\pm q \rangle$ this implies that the matrices $\Phi_{n'n}^\Delta$ and $\Phi_{n'n}^{DE}$ are banded matrices with nonzero entries only for $n'=n, n \pm 2$. Using
\begin{align}
\frac{\pi}{\sin \pi \lambda} &= \frac{1}{\lambda} + \frac{\pi^2}{6} \lambda + \ldots \\
\frac{\pi^2}{\sin^2 \pi \lambda} &= \frac{1}{\lambda^2} + \frac{\pi^2}{3} + \ldots 
\end{align}
we write explicitly
\begin{align}
{\cal T}_\pm[w^n] &= t_{n}^\pm w^{n+2} + r_n^\pm w^n + \ldots 
\end{align}
with 
\begin{align}
t_{n}^\pm &= \frac{1}{2} + \frac{\pm [(-1)^n - 1] - 1}{(n+1) (n+2)} , \\
r_n^\pm &= 
\pi^2 \biggl(\dfrac{1}{6} \pm \frac{1}{3}[(-1)^n - 1] + \dfrac{1\pm 2}{6} \delta_{n1} - \dfrac{1}{6} \delta_{n0}\biggr) \nonumber \\
&= \begin{cases}
0 & n = 0 \\
\dfrac{\pi^2}{3} (1 \mp 1) & n = 1 \\
\pi^2 \biggl(\dfrac{1}{6} \pm \dfrac{1}{3}[(-1)^n - 1]\biggr) & n \geq 2 ,
\end{cases}
\end{align}
which follows from a Taylor expansion of the exponential in the generating function. We then write
\begin{align}
{\cal T}_\pm [u_n] &= \alpha_n^\pm u_{n+2} + \beta_n^\pm u_{n} + \alpha_{n-2}^\pm u_{n-2} ,
\end{align}
where the highest index $n+2$ follows from the fact that the highest monomial in $u_n$ is $w^n$, which can only change to maximal order of $w^{n+2}$, the steps of $2$ follow from the parity-preserving property of ${\cal T}_\pm$, and the absence of lower-order terms $u_{n-4}$ etc.~follows from the Hermitian property and the fact that the $u_n$ are orthogonal polynomials (which implies $\langle u_{n'} | {\cal T}_\pm[u_n] \rangle = \langle {\cal T}_\pm[u_{n'}] | u_n \rangle = 0$ if $n'+2 < n$). Now, use Eq.~\eqref{eq:expansion}, $u_n(w) = \kappa_n (w^n + \nu_n w^{n-2} + \ldots)$, and define
\begin{align}
{\cal T}_\pm [u_n] &= \kappa_n (t_n^\pm w^{n+2} + [r_n^\pm + \nu_n t_{n-2}^\pm] w^{n}) + \ldots \nonumber \\
&= \alpha_n^\pm \kappa_{n+2} w^{n+2} + [\alpha_n^\pm \kappa_{n+2} \nu_{n+2} + \beta_n^\pm \kappa_n] w^{n} + \ldots
\end{align}
This implies
\begin{align}
\alpha_n^\pm &= \frac{\kappa_n}{\kappa_{n+2}} t_n^\pm = \rho_{n+1} \rho_{n+2} t_n^\pm \\
\beta_n^\pm &= r_n^\pm + t_{n-2}^\pm \nu_n - t_n^\pm \nu_{n+2} ,
\end{align}
which gives Eqs.~\eqref{eq:resA} and~\eqref{eq:resC} when substituting the coefficients. The definitions~\eqref{eq:defF} and~\eqref{eq:defH} then give the results~\eqref{eq:resFH}.

Finally, the average $\langle \Sigma_{n'} \Sigma_n \rangle_F$ does not need to be evaluated separately but is inferred from
\begin{align}
&\Delta_{n'} \Delta_n + D_{n'} D_n + E_{n'} E_n + \Sigma_{n'} \Sigma_n \nonumber \\
&= 4 \Bigl( u_{n'}(w_{1}) u_n(w_{1}) + u_{n'}(w_{2}) u_n(w_{2}) \nonumber \\
&\quad + u_{n'}(w_{1'}) u_n(w_{1'}) + u_{n'}(w_{2'}) u_n(w_{2'})\Bigr) .
\end{align}
Hence
\begin{align}
&\Phi_{n'n}^\Delta + 2 \Phi_{n'n}^{DE} + \Phi_{n'n}^\Sigma = 4 J_{n'n} 
\end{align}
with
\begin{align}
J_{n'n} &= \frac{1}{\pi} \biggl\langle u_{n'}(w_1) \biggl| \frac{\pi^2 + w_1^2}{2} \, u_n(w_1) \biggr\rangle = \frac{\rho_n \rho_{n-1}}{2\pi} \delta_{n',n-2}\nonumber \\
&\quad + \frac{\pi^2 + \rho_n^2 + \rho_{n+1}^2}{2\pi} \delta_{n',n} + \frac{\rho_{n+1} \rho_{n+2}}{2\pi} \delta_{n',n+2} .
\end{align}
This matrix is in fact the same (up to normalization) as the matrix $\Gamma_{nn'}$ for a three-dimensional Fermi liquid~\cite{repplinger25,taillat25,taillat26,kurkjian26}. This also eliminates the average $\Phi_{n'n}^\Sigma$ that enters the even-$m$ finite part in Eq.~\eqref{eq:defB} since
\begin{align}
2 \Phi_{n'n}^{DE} + \Phi_{n'n}^\Sigma - 3 \Phi_{n'n}^\Delta &= 4 J_{n'n} - 4 \Phi_{n'n}^\Delta .
\end{align}

\bibliography{bib_oddmodes}

\end{document}